\documentclass[12pt]{article}
\usepackage{epsf}
\begin{document}
\thispagestyle{empty}

\renewcommand {\thefootnote}{\fnsymbol{footnote}}
\renewcommand{\thesection}{\Roman{section}}
\def \beq {\begin{equation}}
\def \eeq {\end{equation}}
\def \bes {\begin{eqnarray}}
\def \ees {\end{eqnarray}}
\def\ni{\noindent}
\def\nn{\nonumber}
\def\rv{\mbox{\boldmath$r$}}
\def\kv{\mbox{\boldmath$k$}}
\def\drv{{d\mbox{\boldmath$r$}}}
\def\mum {$\,\mu\mbox{m}$}
\thispagestyle{empty}
\large
\begin{center}
 { {\bf
{Investigation of the temperature dependence \\of the
Casimir force between real metals}
}}
\vskip 1.5cm
{
G.~L.~Klimchitskaya\footnote{On leave from
North-West Polytechnical Institute,
St.Petersburg, Russia.
Electronic address: galina@fisica.ufpb.br}
and V.~M.~Mostepanenko\footnote{On
leave from
A.Friedmann Laboratory for Theoretical Physics,
St.Petersburg, Russia. Electronic address:
mostep@fisica.ufpb.br}
}
\\[8mm]
{\it
{ Physics Department, Federal University
of Para\'{\i}ba, C.P.5008, \\
CEP 58059--970, Jo\~{a}o Pessoa, Pb---Brazil}
}
\end{center}

\begin{abstract}
We investigate the Casimir force acting between real
metals at nonzero temperature. It is shown that the
zero-frequency term of Lifshitz formula has
interpretation problem in the case of real metal
described by Drude model. It happens because the scattering
theory underlying Lifshitz formula is not well formulated
when the dielectric permittivity takes account of dissipation.
To give the zeroth term of Lifshitz formula the definite
meaning different prescriptions were used recently by
different authors with diversed results. These results
are shown to be improper and in disagreement with
experiment and the general physical requirements.
We propose the new prescription which is a generalization
of Schwinger, DeRaad and Milton recipe formulated earlier
for ideal metals. On this base the detailed numerical
and analytical computations of the temperature Casimir 
force are performed in configuration of two plane
plates and a spherical lens (sphere) above a plate.
The corrections due to nonzero temperature and finite 
conductivity found in the paper are in agreement with the
limiting case of perfect metal and fit all experimental
and theoretical requirements. Among other facts, the
previous results obtained in frames of plasma model
are confirmed. It appears they are the limiting case
of Drude model computations when the relaxation parameter
goes to zero. The comparison with the Casimir force
acting between dielectric test bodies is made.
\end{abstract}

\vskip 8mm
PACS: 12.20.Ds, 11.10.Wx, 12.20.Fv

\section{INTRODUCTION}

The H.B.G.~Casimir effect \cite{1} predicted more than fifty years
ago is one of the most interesting manifestations of zero-point vacuum
oscillations of quantized fields. The Casimir effect implies that there
is some force acting between two uncharged bodies closely spaced in the
vacuum. This effect is purely of quantum origins. There is no such
force in
classical physics. Unique to the Casimir force is its strong dependence
on shape, switching from attractive to repulsive as a function of the
size, geometry and topology of the boundary. The force results from
the alteration by the boundaries of the zero-point electromagnetic
energy that pervades all of space as predicted by quantum field theory.
Alternatively, the Casimir force can be described as the retarded
electromagnetic interaction of atomic and molecular dipoles and was
extended to forces between macroscopic dielectric bodies
characterized by some dielectric constant \cite{2}.

In recent years the Casimir effect has attracted much attention because
of numerous applications in quantum field theory, atomic physics,
condensed matter physics, gravitation and cosmology, and mathematical
physics (see monographs \cite{3,4,5,6} and references therein).
New precision experiments have been performed on measuring the Casimir
force between metallic surfaces \cite{7,8,9,10,11,11a}.
In Refs.~\cite{12,13,14} some promising applications of the Casimir effect
were proposed for diagnostic in thin films and in nano electro-mechanical
systems. Given the above reasons it is very important to understand the
Casimir force between real materials including the effect of such influential
factors as surface roughness, finite conductivity, and nonzero
temperature.

Finite conductivity corrections to the Casimir force have long been
investigated. They were calculated using plasma model
\cite{15,16,17} up to the first order in the relative penetration depth
of the zero-point electromagnetic oscillations into the metal.
In \cite{18} more exact results up to the second order were obtained,
and in \cite{19} --- up to the fourth order. In Refs.~\cite{20,21}
the finite conductivity corrections to the Casimir force were computed
using tabulated optical data for the frequency-dependent complex
refractive index. For all separations between the test bodies larger
than the effective plasma wavelength of the metal under study the results
of \cite{19} and \cite{20,21} are shown to be in good agreement.
The effects of surface roughness in combination with finite conductivity
were investigated in detail in \cite{22} for the configuration of a sphere
above a plate used in experiments \cite{8,9,10,11}. Roughness
contributions to the Casimir effect (including origination of a lateral
force) has been treated recently in \cite{22a,22b,22c,22d}.

The action of nonzero temperature on the Casimir force between dielectric
semispaces was taken into account in the Lifshitz theory \cite{2,15}.
For perfect metals at nonzero temperature the Casimir force was
calculated in \cite{23,24} within the limits of quantum field theory
in terms of the free energy density of vacuum. There were apparent
differences between the results of \cite{2,15} and \cite{23,24} which were
resolved
in \cite{17}. As shown by Schwinger, DeRaad and Milton
\cite{17}, to obtain the case of perfect
conductor from the Lifshitz theory one must take the limit of infinite
dielectric permittivity before putting the frequency equal to zero in
the temperature sum. The results of \cite{2,15} adapted for the case of
perfect metal and of \cite{23,24} are then in agreement.
The temperature corrections to the
Casimir force turned out to be negligible in the experiments \cite{8,9,10,11}
where the measurements were performed in the separation range
$a<1\,\mu$m. However, at $a>1\,\mu$m, as in \cite{7}, the temperature
corrections make large contributions to the zero-temperature force
between perfect conductors (e.g., for $a=5\,\mu$m the temperature
correction in configuration of a spherical lens above a plate exceeds
the zero-temperature force \cite{25}).

The increased accuracy of the Casimir force measurements invites further
investigation of the temperature corrections in case of real metals.
Although from a conceptual point of view the Lifshitz theory provides
a way of obtaining all the required results, the problem here is worse
than it was with the case of perfect metals. In \cite{26} it was suggested
to use
the plasma model in order to describe the dielectric permittivity
along the imaginary frequency axis in the Lifshitz formula for the Casimir
force at nonzero temperature (note that in \cite{15,16,17,18,19} the
plasma model was applied for this purpose at zero temperature only).
In Refs.~\cite{27,28} the detailed calculations of the temperature
Casimir force were performed in framework of the Lifshitz theory and the
plasma
model. It was shown that the temperature corrections are negligible at
small separations where the finite conductivity corrections are very
important. By contrast, at large separations finite conductivity
corrections can be ignored, whereas the temperature corrections play
an important role. In \cite{27,28} the transition region between these
two asymptotic regimes was also investigated where the combined effect of
nonzero temperature and finite conductivity is important and should
be taken into account.

It is a common knowledge that at small frequencies the dielectric
permittivity $\varepsilon$ is proportional to $\omega^{-1}$
as is given by the Drude model. Because of this, the Drude model is favoured
over the plasma model (which implies $\varepsilon\sim\omega^{-2}$)
when calculating the Casimir force at nonzero temperature. The first attempts
to calculate the nonzero temperature Casimir force between real metals
based on
the Lifshitz formula and Drude model were undertaken in \cite{28,29,30,31,32}.
They have led to distinct and unexpected results. It was found that the
value of the transverse reflection coefficient $r_2$ of the electromagnetic
oscillations at zero frequency becomes indefinite when one describes the
boundary made of real metal using the Drude model. In \cite{29,30} the
value $r_2=0$ has been adopted. This assumption leads to a nonphysical
conclusion that the asymptotic Casimir force at high temperature in the case
of real metals is two times smaller than for the case of ideal metal
(without regard to the particular value of the conductivity of real metal).
Also, as a result of assumption, made in \cite{29,30}, there arise large
negative temperature corrections at small separations which are linear in
temperature. These corrections are not only unacceptable from the
theoretical point of view but are in a conflict with experimental data
\cite{33}.

The other authors \cite{31,32} assumed the value $|r_2|=1$ 
at zero frequency using the
prescription of \cite{17} formulated for perfect metal and well known
relation by Hagen and Rubens which is valid for real photons only
\cite{34,35}. This assumption also leads to nonphysical conclusions,
i.e. to linear (although positive) temperature corrections at small
separations and to the absence of any finite conductivity corrections
to the Casimir force for real metals starting from the moderate
separations of several micrometers regardless of metal quality
(note that the same assumption was accepted in the latest version
of \cite{33}).

The situation was clarified in \cite{28} where the discontinuity of the
transverse reflection coefficient as a function of frequency and photon
momentum was demonstrated in the case of real metal described by the
Drude model. According to \cite{28} to clear away the ambiguity in the
zero frequency term of Lifshitz formula arising due to this discontinuity
it is necessary to use an alternative representation for it \cite{17,23}.
This representation gives the possibility to redefine the zero-frequency
term of Lifshitz formula in order to assign it the definite meaning
for real metals in accordance with the usual physical requirements.
In \cite{28}, however, the values of the Casimir force including 
nonzero temperature and finite conductivity in
the framework of the Drude model were not computed. Thus, up to now, there
is no
plausible qualitative information on the Casimir effect between real
metals including the temperature corrections. The need for such
investigation is apparent when the experimental and technological
applications of the Casimir force mentioned above are considered.
It is also important that recently, the Casimir effect has
been used to obtain stronger constraints on the constants of long-range
interactions (including corrections to the Newtonian gravitational law)
predicted by the unified field theories, supersymmetry, supergravity
and string theory \cite{25,36,37,38,39}. The reliable theoretical values
of the Casimir force at nonzero temperature between real metals and
the extent of their agreement with experiment are of particular interest
in order to obtain the strongest constraints.

In the present paper we propose a method which allows one to attribute
a definite value to the term of Lifshitz formula at zero
frequency for real metals as described by the Drude model. This method 
avoids the above mentioned contradictions and solves the problem
in a physically consistent way. The detailed computations of the Casimir
force between real metals at nonzero temperature are performed for the
configuration of two plane parallel plates and a sphere (lens) above
a plate in a wide separation range from $0.1\,\mu$m to $10\,\mu$m.
It is shown that at small separations (low temperatures) the
temperature corrections are small irrespective of whether the Drude
or plasma model is used. In particular, no corrections to the
force arise which would be linear in temperature, such as  in
\cite{29,30,31,32}.
At large separations (high temperatures) there is some difference between
the finite conductivity corrections to the temperature Casimir force
obtained with the plasma and Drude models, although the corrections themselves
are rather small. With decreasing relaxation parameter the results from 
both models coincide. Our results at large separations join smoothly
with the increase of conductivity with the asymptotic values obtained for
the perfect metal (this is not the case in \cite{29,30} where the asymptotic
force for a real metal is two times smaller than for the perfect one).
Also the non-physical results of \cite{31,32},  according to which
at large separations the finite conductivity corrections are absent in the
case of real metals, are shown to be in error. Below the nonzero 
temperature Casimir force between the
dielectric bodies is also computed  and the distinctions between the
cases of metallic and dielectric bodies are discussed. The obtained
results are in agreement with the performed experiments. They establish
the theoretical basis for new precise experiments on measuring the
Casimir force.

The paper is organized as follows. In Sec.~II the general formalism for
the Casimir force between real metals at nonzero temperature is presented
for the configuration of two plane parallel plates. Here special attention
is paid to the indefinite character and discontinuity of the zeroth term
of the Lifshitz formula in case of the Drude model. The representation of this
term is given which is in accordance with the general physical
requirements. In Sec.~III the same is done for the configuration of a sphere
(lens) above a plate. Sec.~IV is devoted to the numerical and analytical
computations of the temperature Casimir force for the configuration
of two plane parallel plates. Both the asymptotics of low and high
temperatures are considered and also a transition region between them.
Sec.~V contains the results of analogous computations for the configuration
of a sphere (lens) above a plate. In Sec.~VI the case of dielectrical
test bodies is considered and the Casimir force at nonzero temperature
is found. In Sec.~VII reader will find conclusions and discussion. 
Appendices I and II
contain some details of the mathematical calculations.

\section{TEMPERATURE CASIMIR FORCE BETWEEN TWO PLANE
PARALLEL PLATES}

The original Lifshitz derivation \cite{2,15} of the temperature Casimir
force between two semispaces was based on the assumption that the
dielectric materials can be considered as continuous media characterized
by randomly fluctuating sources. The correlation function of these
sources, situated at different points, is proportional to the
$\delta$-function of the radius-vector joining these points. The force
per unit area acting upon one of the semispaces was calculated as
the flux of incoming momentum into the semispace through the boundary
plane. This flux is given by the appropriate component of the stress
tensor ($zz$-component if $xy$ is the boundary plane). Usual boundary
conditions on the boundary surfaces between different media were imposed
on the temperature Green's functions. To exclude the divergences, the
values of all the Green's functions in vacuum were subtracted from their
values in dielectric media.
The other derivation of Lifshitz formula is based on the solution of
Maxwell's equations with appropriate boundary conditions on the surfaces
separating different media \cite{40,41,42}. 
In this manner the allowed surface
modes and a harmonic oscillator free energy of each mode can be
determined. The total renormalized free energy is obtained by the
summation over all modes and subtraction of the free energy of  vacuum.
The nonzero temperature Casimir force is finally given by the negative
derivative
of the renormalized free energy with respect to the distance between
semispaces (at zero temperature this procedure is presented in detail in
\cite{21,42}).

The modern derivation of Lifshitz formula \cite{43}
is based on the temperature
field theory in Matsubara formulation. In case of a static Casimir effect
the system is in thermal equilibrium. To describe it at nonzero
temperature one should take the Euclidean version of the field theory
with a field periodic in the time variable within  the interval
$\beta=\hbar/(k_{B}T)$, where $T$ is temperature, $k_{B}$ is the
Boltzmann constant. To find the free energy in case of two semispaces
the scattering problem on the $z$-axis perpendicular to the boundary
planes is considered. An electromagnetic wave coming from the left
in the dielectric is scattered on the air gap and there exist a transmitted
and a reflected wave. Finally the free energy and the Casimir force is
expressed in terms of the scattering coefficient on the imaginary axis.
Calculating this coefficient for the problem of two semispaces 
(notated by an index $ss$) with
a frequency dependent dielectric permittivity $\varepsilon(\omega)$
separated by a gap of width $a$ one finally obtains the Casimir force
in the form \cite{43}
\bes
&&
F_{ss}(a)=-\frac{k_BT}{2\pi}
\sum\limits_{l=-\infty}^{\infty}
\int\limits_{0}^{\infty}k_{\bot}\,dk_{\bot}\,q_l
\left\{\left[r_1^{-2}(\xi_l,k_{\bot})e^{2aq_l}-1\right]^{-1}\right.
\nn \\
&&\phantom{aaaaaaaaaaaaaaaaaaaaa}
+\left.\left[r_2^{-2}(\xi_l,k_{\bot})e^{2aq_l}-1\right]^{-1}\right\},
\label{1}
\ees
\ni
where $r_{1,2}$ are the reflection coefficients with parallel
(perpendicular) polarization respectively given by
\beq
r_1^{-2}(\xi_l,k_{\bot})=\left[
\frac{\varepsilon(i\xi_l)q_l+k_l}{\varepsilon(i\xi_l)q_l-k_l}\right]^2,
\qquad
r_2^{-2}(\xi_l,k_{\bot})=\left(
\frac{q_l+k_l}{q_l-k_l}\right)^2.
\label{2}
\eeq
\ni
Here ${\kv}_{\bot}$ is the momentum component lying in the boundary planes,
$k_{\bot}=|{\kv}_{\bot}|$, $\omega=i\xi$, and the following notations
are used
\beq
q_l=\sqrt{\frac{\xi_l^2}{c^2}+k_{\bot}^2}, \quad
k_l=\sqrt{\varepsilon(i\xi_l)\frac{\xi_l^2}{c^2}+k_{\bot}^2},
\quad
\xi_l=\frac{2\pi l}{\beta}.
\label{3}
\eeq

Introducing a new variable $p$ according to
\beq
k_{\bot}^2=\frac{\xi_l^2}{c^2}(p^2-1)
\label{4}
\eeq
we rewrite Eq.~(\ref{1}) in the original Lifshitz form \cite{2,15,44}
\bes
&&
F_{ss}(a)=-\frac{k_BT}{\pi c^3}
\sum\limits_{l=0}^{\infty}{\vphantom{\sum}}^{\prime}\xi_l^3
\int\limits_{1}^{\infty}p^2\,dp\,
\left\{\left[\left(
\frac{K(i\xi_l)+\varepsilon(i\xi_l)p}{K(i\xi_l)-\varepsilon(i\xi_l)p}
\right)^2e^{2a\frac{\xi_l}{c}p}-1\right]^{-1}\right.
\nn\\
&&
\phantom{F_{ss}^{T}(a)}+\left.
\left[\left(
\frac{K(i\xi_l)+p}{K(i\xi_l)-p}
\right)^2e^{2a\frac{\xi_l}{c}p}-1\right]^{-1}\right\},
\label{5}
\ees
\ni
where 
\beq
K(i\xi_l)\equiv\left[p^2-1+\varepsilon(i\xi_l)\right]^{1/2}
\label{6}
\eeq
\ni
and the
prime near the 
summation sign means that the zeroth term is taken with the
coefficient 1/2.

Note that the representation of Eq.~(\ref{5}) for the nonzero temperature
Casimir force has a disadvantage as the $l=0$ term in it is the product
of zero by a divergent integral. This is usually eliminated \cite{44}
by introducing the variable $y$ instead of $p$
\beq
y=\left\vert{\tilde\xi}_l\right\vert p, \qquad
{\tilde\xi}_l=2a\frac{\xi_l}{c},
\label{7}
\eeq
\ni
where ${\tilde\xi}_l$ is the dimensionless frequency. In terms of new
variables Eq.~(\ref{5}) is
\bes
&&
F_{ss}(a)=-\frac{k_BT}{16\pi a^3}
\sum\limits_{l=-\infty}^{\infty}
\int\limits_{|{\tilde\xi}_l|}^{\infty}y^2dy
\left\{\left[r_1^{-2}({\tilde\xi}_l,y)e^{y}-1\right]^{-1}\right.
\nn \\
&&\phantom{aaaaaaaaaaaaaaaaaaaaa}
+\left.\left[r_2^{-2}({\tilde\xi}_l,y)e^{y}-1\right]^{-1}\right\},
\label{8}
\ees
\ni
where
\bes
&&
r_1({\tilde\xi}_l,y)=
\frac{\varepsilon y-\sqrt{(\varepsilon -1){\tilde\xi}_l^2+
y^2}}{\varepsilon y+\sqrt{(\varepsilon -1){\tilde\xi}_l^2+y^2}},
\label{9}\\
&&
r_2({\tilde\xi}_l,y)=
\frac{y-\sqrt{(\varepsilon -1){\tilde\xi}_l^2+
y^2}}{y+\sqrt{(\varepsilon -1){\tilde\xi}_l^2+y^2}},
\quad
\varepsilon\equiv\varepsilon(i\xi_l)=
\varepsilon\left(i\frac{c{\tilde\xi}_l}{2a}\right).
\nn
\ees
\ni
Both changes of variables (\ref{4}) and (\ref{7}) are, however,
singular at $l=0$. In fact (\ref{8}) can be obtained from (\ref{1})
by the regular change of variable
\beq
4a^2k_{\bot}^2=y^2-{\tilde\xi}_l^2.
\label{10}
\eeq
\ni
Because of this, the equivalent representations of Lifshitz formula
(\ref{1}) and (\ref{8}) are preferred as compared to (\ref{5}).

To calculate the Casimir force at nonzero temperature between real metals
one should substitute the appropriate values $\varepsilon(i\xi_l)$ into
Eq.~(\ref{1}) or (\ref{8}). In some frequency range $\varepsilon(i\xi_l)$
can be found by the use of optical tabulated data for the complex
refractive index for the metal under consideration tabulated in, e.g.,
\cite{45} (this was done in \cite{20,21}). But in any case the optical
data should be extrapolated outside the region where they are available
in the tables to smaller and larger frequencies. This can be performed
by the use of plasma model function
\beq
\varepsilon_p(i\xi)=1+\frac{\omega_p^2}{\xi^2}
\label{11}
\eeq
\ni
or more exact Drude model one
\beq
\varepsilon_D(i\xi)=1+\frac{\omega_p^2}{\xi(\xi+\gamma)},
\label{12}
\eeq
\ni
where $\omega_p$ is the plasma frequency and $\gamma$ is the relaxation
frequency. For some metals, the model dielectric functions with
appropriate values of  $\omega_p$ and $\gamma$ can be reliably used
throughout the whole spectrum. Calculations of this type were performed
in \cite{27,28,29,30,31,32} with different results. As shown below, the
reason for this is the discontinuity of the transverse reflection
coefficient $r_2$ at zero frequency with respect to
relaxation parameter
which must be made continuous in a physically
reasonable way in order to describe real metals by the Lifshitz formula.

If we put $\xi_l=0$ in Eqs.~(\ref{2}), (\ref{3}) and use the plasma
model dielectric function (\ref{11}) the result is \cite{27,28}
\beq
r_1^2(0,k_{\bot})=1, \quad
r_2^2(0,k_{\bot})=\left(\frac{k_{\bot}-\sqrt{k_{\bot}^2+
\frac{\omega_p^2}{c^2}}}{k_{\bot}+\sqrt{k_{\bot}^2+
\frac{\omega_p^2}{c^2}}}\right)^2.
\label{13}
\eeq
\ni
In the case of the Drude model dielectric function (\ref{12}) it holds
\beq
\left.\varepsilon_D(i\xi_l)\frac{\xi_l^2}{c^2}\right\vert_{l=0}=0,
\label{14}
\eeq
\ni
which leads to \cite{30}
\beq
r_1^2(0,k_{\bot})=1, \quad
r_2^2(0,k_{\bot})=0
\label{15}
\eeq
\ni
for any $k_{\bot}\neq 0$. Note that (\ref{15}) is valid for arbitrary small
$\gamma$ and arbitrary large $\omega_p$. Thus the second equation from
(\ref{15}) is in contradiction with the limiting case of ideal metal
$r_2^2(0,k_{\bot})=1$ which follows from (\ref{13}) in the limit
$\omega_p\to\infty$. The results obtained using (\ref{15}) are in
conflict with the known results for ideal metal \cite{17,24}. What is more
important, the second equation of (\ref{15}) does not approach the
second equation of (\ref{13}) when the relaxation frequency $\gamma$
goes to zero. In the limit $\gamma\to 0$ one still has
$r_2^2(0,k_{\bot})=0$, not (\ref{13}), although
$\lim\limits_{\gamma\to 0}\varepsilon_D=\varepsilon_p$ in accordance
with Eqs.~(\ref{11}), (\ref{12}). In Secs.~IV,\,V the discontinuity of the
zero frequency term of the Lifshitz formula with respect to the relaxation
parameter will be discussed in detail when comparing the continuous
modification of this term as suggested below. Exactly the same results,
as in (\ref{13}), (\ref{15}), are obtained if one uses the variables
(${\tilde\xi},\,y$) instead of ($\xi,\,k_{\bot}$) and 
the representation (\ref{8}) of the Lifshitz formula.

If the original representation (\ref{5}) of the Lifshitz formula is exploited
and written in terms of the variables ($\xi,\,p$) the situation changes
drastically. Here one immediately arrives at
\beq
r_1^2(0,p)=r_2^2(0,p)=1
\label{16}
\eeq
\ni
for all $p\neq\infty$ irrespective of whether the plasma or Drude model is
used for the dielectric permittivity on the imaginary axis. The reason for
the distinct value of $r_2(0,p)$ is the singular character of the change of
variables (\ref{4}) when $\xi_0=0$. This change relates the single
value of $k_{\bot}=0$ to all the finite values of $p$, and all the values
of  $k_{\bot}\neq 0$ to a single point $p=\infty$. The values (\ref{16})
for both reflection coefficients at zero frequency were postulated
at \cite{31,32} and used in all the numerical computations
even after one more singular change of variables (\ref{7}) was performed
in order to calculate the divergent integral in Eq.~(\ref{5}). We show
below, that this postulate, though it works good for the perfect metal,
is not justified for real metals of finite conductivity.

In actual truth, the transverse reflection coefficient $r_2$ from
Eq.~(\ref{2}) or (\ref{9}) has a discontinuity as a function of two continuous
variables ($\tilde\xi$ and $y$ for instance) in the case of the Drude
dielectric function \cite{28}. If one puts $\tilde\xi=0$ in (\ref{9})
from the very beginning the result is $r_2^2(0,y)=0$ in agreement with
(\ref{15}). If, however, one approaches the point ($\tilde\xi=0,\,y=0$)
along the direction $\tilde\xi=ky$ in the ($\tilde\xi,y$)-plane
then $r_2^2(\tilde\xi,y)\to 1$.  The distinguishing feature of the Drude
model is that for an arbitrarily small $y$ there exists $k$ such that
$r_2^2(ky,y)$ takes any value in between zero and unity. By contrast, in the
case of plasma model both reflection coefficients are continuous, and
$r_2^2(ky,y)$ does not depend on $k$ but is determined only by the value
of $y$. Of even greater concern is that the transverse reflection
coefficient at zero frequency is discontinuous with respect to the relaxation
parameter $\gamma$.

As is seen from the above discussion, there is a serious unresolved issue
concerning the value of the zero frequency term of the Lifshitz formula for
real metals. In fact the scattering problem, which forms the basis of the
Lifshitz formula, is meaningful only 
for nondissipative media  (this is the case for plasma model or for dielectric
materials). The case $\varepsilon=\infty$, as it is for metals at zero
frequency, is especially complicated when the Drude model is used taking
account of dissipation. The latter leads to the violation of the unitarity
condition and thereby the zeroth term of Lifshitz formula becomes indefinite
and must be redefined.
For ideal metals of infinite conductivity this issue was
resolved by the Schwinger, DeRaad, and Milton prescription \cite{17}
demanding that the limit $\varepsilon\to\infty$ should be taken before
setting $\xi=0$ (which is equivalent to the use of Eq.~(\ref{16})).
Let us find out, in what way the problem of zero-frequency term of
the Lifshitz formula can be solved for real Drude metals in a physically
satisfactory way.

For this purpose let us use the representation of the Lifshitz formula in
terms
of continuous frequency variable. Such a representation was suggested in
\cite{23} and used in \cite{17,24} for the other purposes.
According to the Poisson summation formula if $c(\alpha)$ is the Fourier
transform of a function $b(x)$
\beq
c(\alpha)=\frac{1}{2\pi}
\int\limits_{-\infty}^{\infty}b(x)e^{-i\alpha x}\,dx
\label{17}
\eeq
\noindent
then it follows
\beq
\sum\limits_{l=-\infty}^{\infty}b(l)=
2\pi\sum\limits_{l=-\infty}^{\infty}c(2\pi l).
\label{18}
\eeq

We apply this formula to Eq.~(\ref{8}) using the
identification
\beq
b_{ss}(l)\equiv -\frac{k_BT}{16\pi a^3}
\int\limits_{|l|\tau}^{\infty}y^2\,dy\,f_{ss}(l\tau,y),
\qquad
\tau\equiv\frac{4\pi ak_BT}{\hbar c},
\label{19}
\eeq
\noindent
where ${\tilde\xi}_l=\tau l$ and
\bes
&&
f_{ss}(l\tau,y)=f_{ss}^{(1)}(l\tau,y)+f_{ss}^{(2)}(l\tau,y),
\label{20}\\
&&
f_{ss}^{(i)}(l\tau,y)=\left(r_i^{-2}e^y-1\right)^{-1},
\quad i=1,\,2
\nonumber
\ees
\noindent
are the even functions of $l$.

Then the quantity $c_{ss}(\alpha)$ from Eq.~(\ref{17})
is given by
\beq
c_{ss}(\alpha)=-\frac{k_BT}{16\pi^2 a^3}
\int\limits_{0}^{\infty}dx\,\cos\alpha x
\int\limits_{x\tau}^{\infty}y^2\,dy\,f_{ss}(x\tau,y).
\label{21}
\eeq

Using Eqs.~(\ref{8}),  (\ref{18}), (\ref{21}) one
finally obtains the representation of the Lifshitz formula
\bes
&&
F_{ss}(a)=
\sum\limits_{l=-\infty}^{\infty}b_{ss}(l)
\label{22}\\
&&\phantom{aaa}
=-\frac{\hbar c}{16\pi^2 a^4}
\sum\limits_{l=0}^{\infty}{\vphantom{\sum}}^{\prime}
\int\limits_{0}^{\infty}d{\tilde\xi}\,
\cos\left(l{\tilde\xi}\frac{T_{eff}}{T}\right)
\int\limits_{\tilde\xi}^{\infty}y^2\,dy\,f_{ss}(\tilde\xi,y),
\nonumber
\ees
\ni
where the continuous frequency variable $\tilde\xi=\tau x$,
and $k_BT_{eff}=\hbar c/(2a)$.
Note that in the representation (\ref{22}) the $l=0$ term gives the
force at zero temperature. It is useful also to change the order of
integration  in Eq.~(\ref{22})
\beq
F_{ss}(a)=
-\frac{\hbar c}{16\pi^2 a^4}
\sum\limits_{l=0}^{\infty}{\vphantom{\sum}}^{\prime}
\int\limits_{0}^{\infty}y^2\,dy
\int\limits_{0}^{y}d{\tilde\xi}
\cos\left(l{\tilde\xi}t\right)
f_{ss}(\tilde\xi,y),
\label{23}
\eeq
\ni
where $t\equiv T_{eff}/T$.

Let us isolate the zero-frequency term of the usual Lifshitz formula
(\ref{8}) with discrete frequencies in representation (\ref{23}). In this
a way it will be expressed in terms of the integrals with respect to
continuous variables $\tilde\xi,\,y$. For this purpose we write out
separately the terms of Eq.~(\ref{23}) with $l=0$, integrate all the
other terms by parts with respect to $\tilde\xi$, change the order
of summation and integrations, and use the formula \cite{46}
\beq
\sum\limits_{l=1}^{\infty}\frac{\sin(lzt)}{l}=\frac{1}{2}
\left[\pi+2\pi A\left(\frac{tz}{2\pi}\right)-tz\right],
\label{24}
\eeq
\ni
where $A(z)$ is the integer portion of $z$. The result of these
transformations is
\bes
&&
F_{ss}(a)=
-\frac{\hbar c}{32\pi^2 a^4}\left\{
\int\limits_{0}^{\infty}y^2\,dy
\int\limits_{0}^{y}d{\tilde\xi}
f_{ss}(\tilde\xi,y)\right.
\label{25}\\
&&\phantom{F_{ss}(a)}
+\frac{1}{t}
\int\limits_{0}^{\infty}y^2\,dy\,f_{ss}(y,y)
\left[\pi+2\pi A\left(\frac{ty}{2\pi}\right)-ty\right]
\nn \\
&&\phantom{F_{ss}(a)}
-\frac{1}{t}\left.
\int\limits_{0}^{\infty}y^2\,dy
\int\limits_{0}^{y}d{\tilde\xi}
\frac{\partial f_{ss}(\tilde\xi,y)}{\partial{\tilde\xi}}
\left[\pi+2\pi A\left(\frac{t{\tilde\xi}}{2\pi}\right)-
t{\tilde\xi}\right]\right\}.
\nn
\ees
\ni
Here the first term is the zero-temperature force. The second and the third
contributions in the right-hand side of Eq.~(\ref{25}) can be transformed
using the definition of the function $A(z)$ and representation of
$f_{ss}$ in terms of the sum of parallel and transverse modes in
accordance with (\ref{20}). Taking into account that $f_{ss}^{(1)}$ and
its derivative are continuous functions (as distinct from $f_{ss}^{(2)}$)
one obtains finally after the cancellation of 
zero-temperature
contribution (see Appendix I for details)
\beq
F_{ss}(a)=F_{ss}^{(l=0)}(a)
-\frac{k_BT}{8\pi a^3}
\sum\limits_{l=1}^{\infty}
\int\limits_{{\tilde\xi}_l}^{\infty}y^2dy
f_{ss}({\tilde\xi}_l,y).
\label{26}
\eeq
\ni
In this equation all the terms with $l\geq 1$ coincide with those in
Eq.~(\ref{8}) and the term with $l=0$ is given by
\bes
&&
F_{ss}^{(l=0)}(a)=-\frac{k_BT}{16\pi a^3}\left\{
\int\limits_{0}^{\infty}y^2dy
\left[f_{ss}^{(1)}(0,y)+f_{ss}^{(2)}(y,y)\right]\right.
\nn \\
&&\phantom{F_{ss}^{(l=0)}(a)}
\left.-
\int\limits_{0}^{\infty}y^2dy
\int\limits_{0}^{y}d{\tilde\xi}
\frac{\partial f_{ss}^{(2)}(\tilde\xi,y)}{\partial{\tilde\xi}}\right\}.
\label{27}
\ees

The obtained representation for the zero-frequency term of the Lifshitz
formula
is well suited for solving the problem formulated above. In terms of
dimensionless variables (${\tilde\xi}, \,y$) the plasma and Drude
dielectric functions take the form
\beq
\varepsilon_p(i{\tilde\xi})=1+\frac{{\tilde\omega}_p^2}{{\tilde\xi}^2},
\quad
\varepsilon_D(i{\tilde\xi})=1+
\frac{{\tilde\omega}_p^2}{{\tilde\xi}({\tilde\xi}+{\tilde\gamma})},
\label{28}
\eeq
\ni
where ${\tilde\omega}_p\equiv 2a\omega_p/c$,
${\tilde\gamma}\equiv 2a\gamma/c$. Evidently, no discontinuity
problem arises in the single integral with respect to $y$ in the
right-hand side of Eq.~(\ref{27}). It is well defined for both
dielectric functions of Eq.~(\ref{28}) and for both polarizations, and
contains the limiting cases of an ideal metal and zero relaxation parameter.
What this means is that the single integral term in (\ref{27}) is analytic
with respect to $\tilde\gamma$, i.e. its values calculated with the
Drude model approach the values calculated with the plasma model when
$\tilde\gamma\to 0$ (see also the numerical computations of Sec.~IV).

A completely different type of situation occurs in the double integral
from the right-hand side of Eq.~(\ref{27}). In fact, by the use of
(\ref{9}), (\ref{20}) one obtains
\beq
\frac{\partial f_{ss}^{(2)}(\tilde\xi,y)}{\partial{\tilde\xi}}=
2\frac{e^y}{\left[e^y-r_2^2(\tilde\xi,y)\right]^2}
r_2(\tilde\xi,y)
\frac{\partial r_2(\tilde\xi,y)}{\partial{\tilde\xi}},
\label{29}
\eeq
\ni
where
\beq
\frac{\partial r_2(\tilde\xi,y)}{\partial{\tilde\xi}}=
-\frac{2y}{\left[\sqrt{(\varepsilon-1){\tilde\xi}^2+y^2}+y\right]^2}
\frac{2{\tilde\xi}(\varepsilon-1)+
{\tilde\xi}^2\frac{d\varepsilon}{d{\tilde\xi}}}{2\sqrt{(\varepsilon-
1){\tilde\xi}^2+y^2}}.
\label{30}
\eeq
\ni
Substitution of $\varepsilon=\varepsilon_p$ from Eq.~(\ref{28}) into
(\ref{29}), (\ref{30}) leads to
\beq
\frac{\partial r_2(\tilde\xi,y)}{\partial{\tilde\xi}}=
\frac{\partial f_{ss}^{(2)}(\tilde\xi,y)}{\partial{\tilde\xi}}=0,
\label{31}
\eeq
\ni
so that in the case of the plasma model the double integral in (\ref{27})
vanishes. If, however, $\varepsilon=\varepsilon_D$ is substituted into
(\ref{29}), (\ref{30}) one discovers a discontinuity in $\tilde\gamma$.
Actually, as it follows from (\ref{29}), (\ref{30})
\beq
\lim\limits_{\tilde\gamma\to 0}
\frac{\partial f_{ss}^{(2)}(\tilde\xi,y)}{\partial{\tilde\xi}}=0
\label{32}
\eeq
\ni
for any $\tilde\xi\geq 0$ and $y\neq 0$. On the other hand,
\beq
\lim\limits_{\tilde\gamma\to 0}
\int\limits_{0}^{y}d{\tilde\xi}
\frac{\partial f_{ss}^{(2)}(\tilde\xi,y)}{\partial{\tilde\xi}}=
\left[\left(\frac{y+\sqrt{{\tilde\omega}_p^2+y^2}}{y-
\sqrt{{\tilde\omega}_p^2+y^2}}\right)^2e^y-1\right]^{-1}\neq 0.
\label{33}
\eeq
\ni
To conclude, the presence of the double integral in the right-hand side
of Eq.~(\ref{27}) renders it unsuitable for real metals described by the
Drude model.

To remedy the situation let us notice that under the increase of
$\varepsilon$ it holds
\beq
\frac{\partial f_{ss}^{(2)}(\tilde\xi,y)}{\partial{\tilde\xi}}
\sim\frac{1}{\sqrt{\varepsilon-1}}{\to}0
\quad \mbox{with}{\ }{\varepsilon\to\infty}.
\label{34}
\eeq
\ni
Consequently, the discontinuous term can be deleted by 
the use of Schwinger, DeRaad and Milton-type 
prescription \cite{17} which is that the limit of
dielectric permittivity to infinity must be performed before setting
frequency to zero when considering the zero-frequency term of the Lifshitz
formula. For ideal metals the use of such a prescription is necessary
in order to achieve agreement between the Lifshitz formula for
dielectrics and field theoretical results obtained for a perfect metal, i.e.
for boundaries with the Dirichlet boundary condition. As it is advocated
here, for real metals the analogous prescription must be 
used in a more restricted way to eliminate 
the discontinuity in the zero-frequency term.

With the above prescription the Casimir force acting between two plane
parallel
plates at nonzero temperature is given by
\bes
&&
F_{ss}(a)=-\frac{k_BT}{16\pi a^3}\left\{
\int\limits_{0}^{\infty}y^2dy
\left[f_{ss}^{(1)}(0,y)+f_{ss}^{(2)}(y,y)\right]\right.
\nn \\
&&\phantom{F_{ss}(a)}
\left.+2
\sum\limits_{l=1}^{\infty}
\int\limits_{{\tilde\xi}_l}^{\infty}y^2dy
f_{ss}({\tilde\xi}_l,y)\right\}.
\label{35}
\ees
\ni
In Sec.~IV the detailed calculations of the Casimir force using Eq.~(\ref{35})
are performed. The obtained results are compared with those obtained
by other authors.

\section{TEMPERATURE CASIMIR FORCE IN CONFIGURATION OF A SPHERICAL
LENS (SPHERE) ABOVE A PLATE}

In this section the most important results of Sec.~II are adapted for
a lens (sphere) above a semispace (plate) which is the configuration
most often used in experiments \cite{7,8,9,10,11}. Although there is
no fundamental derivation of the Casimir force acting between a lens
and a plate the proximity force theorem \cite{47} gives the possibility
to express it in terms of the Casimir free energy density for the
configuration of two plane parallel plates. This latter is given by
\cite{27,28,43}
\bes
&&
E_{ss}(a)=\frac{k_BT}{4\pi}
\sum\limits_{l=-\infty}^{\infty}
\int\limits_{0}^{\infty}k_{\bot}\,dk_{\bot}
\left\{\ln\left[1-r_1^{2}(\xi_l,k_{\bot})e^{-2aq_l}\right]\right.
\nn \\
&&\phantom{aaaaaaaaaaaaaaaaaaaaa}
+\left.\ln\left[1-r_2^{2}(\xi_l,k_{\bot})e^{-2aq_l}\right]\right\},
\label{36}
\ees
\ni
where the reflection coefficients and other notations are introduced
in Eqs.~(\ref{2}), (\ref{3}). Note that the Casimir force between plates
from Eq.~(\ref{1}) can be obtained as
$F_{ss}=-\partial E_{ss}/\partial a$.

According to the proximity force theorem \cite{47} the Casimir force
acting between a lens (sphere) of radius $R$ and a plate is
\bes
&&
F_{sl}(a)=2\pi RE_{ss}(a)
\nn \\
&&\phantom{F_{sl}(a)}
=\frac{k_BTR}{2}
\sum\limits_{l=-\infty}^{\infty}
\int\limits_{0}^{\infty}k_{\bot}\,dk_{\bot}
\left\{\ln\left[1-r_1^{2}(\xi_l,k_{\bot})e^{-2aq_l}\right]\right.
\nn \\
&&\phantom{aaaaaaaaaaaaaaaaaaaaa}
+\left.\ln\left[1-r_2^{2}(\xi_l,k_{\bot})e^{-2aq_l}\right]\right\}.
\label{37}
\ees
\ni
Index $sl$ here stands for semispace-lens.
This formula is valid with rather high accuracy of about $a/R$ \cite{48}
which is usually a fraction of a percent.

Introducing the dimensionless variables ${\tilde\xi}_l$ and $y$ from
Eqs.~(\ref{7}), (\ref{10}) we rewrite Eq.~(\ref{37}) in the form
\bes
&&
F_{sl}(a)=
\frac{k_BTR}{8a^2}
\sum\limits_{l=-\infty}^{\infty}
\int\limits_{|{\tilde\xi}_l|}^{\infty}y\,dy
\left\{\ln\left[1-r_1^{2}({\tilde\xi}_l,y)e^{-y}\right]\right.
\nn \\
&&\phantom{aaaaaaaaaaaaaaaaaaaaa}
+\left.\ln\left[1-r_2^{2}({\tilde\xi}_l,y)e^{-y}\right]\right\}.
\label{38}
\ees
\ni
This result is in direct analogy with Eq.~(\ref{8}) for two plates.
There is no need to use singular changes of variables as discussed
in the previous section.

The term of Eq.~(\ref{38}) with $l=0$ suffers exactly the same ambiguity
as the zeroth term of Eq.~(\ref{8}) if one uses the Drude model (\ref{12})
to describe the dependence of a dielectric permittivity on frequency.
This ambiguity lies in the fact that the quantity
$r_2^2(\tilde\xi,y)$ has a discontinuity at a point (0,0) as a function of two
variables and is discontinuous also at the point of zero relaxation
parameter (see the detailed discussion in Sec.~II). Because of this,
the zeroth term of Eq.~(\ref{38}), when applied to real metals,
must be modified in the same way as it was done for configuration of
two plane plates. For this purpose we apply the Poisson summation
formula (\ref{18}) and rewrite Eq.~(\ref{38}) in the form analogical to
Eq.~(\ref{22})
\beq
F_{sl}(a)=
\frac{\hbar cR}{8\pi a^3}
\sum\limits_{l=0}^{\infty}{\vphantom{\sum}}^{\prime}
\int\limits_{0}^{\infty}d{\tilde\xi}\,
\cos\left(l{\tilde\xi}\frac{T_{eff}}{T}\right)
\int\limits_{\tilde\xi}^{\infty}y\,dy\,f_{sl}(\tilde\xi,y),
\label{39}
\eeq
\ni
where
\bes
&&
f_{sl}(\tilde\xi,y)=f_{sl}^{(1)}(\tilde\xi,y)+
f_{sl}^{(2)}(\tilde\xi,y),
\label{40} \\
&&f_{sl}^{(i)}(\tilde\xi,y)=\ln\left(1-r_i^2e^{-y}\right).
\nn
\ees
\ni
Evidently, the zeroth term of Eq.~(\ref{39}) gives us the Casimir force at
zero temperature. The terms with $l\geq 1$ represent the temperature
correction. After changing the order of integration in Eq.~(\ref{39})
one obtains
\beq
F_{sl}(a)=
\frac{\hbar cR}{8\pi a^3}
\sum\limits_{l=0}^{\infty}{\vphantom{\sum}}^{\prime}
\int\limits_{0}^{\infty}y\,dy
\int\limits_{0}^{y}d{\tilde\xi}
\cos\left(l{\tilde\xi}t\right)
f_{sl}(\tilde\xi,y).
\label{41}
\eeq

Using Eq.~(\ref{24}) and performing exactly the same transformations as
in Sec.~II and Appendix I, the Casimir force between a plate and a lens
takes the form
\beq
F_{sl}(a)=F_{sl}^{(l=0)}(a)
+\frac{k_BTR}{4 a^2}
\sum\limits_{l=1}^{\infty}
\int\limits_{{\tilde\xi}_l}^{\infty}y\,dy
f_{sl}({\tilde\xi}_l,y).
\label{42}
\eeq
\ni
Here all terms with $l\geq 1$ are exactly the same as in
Eq.~(\ref{38}), whereas the zeroth term  is given by
\bes
&&
F_{sl}^{(l=0)}(a)=\frac{k_BTR}{8 a^2}\left\{
\int\limits_{0}^{\infty}y\,dy
\left[f_{sl}^{(1)}(0,y)+f_{sl}^{(2)}(y,y)\right]\right.
\nn \\
&&\phantom{F_{ss}^{(l=0)}(a)}
\left.-
\int\limits_{0}^{\infty}y\,dy
\int\limits_{0}^{y}d{\tilde\xi}
\frac{\partial f_{sl}^{(2)}(\tilde\xi,y)}{\partial{\tilde\xi}}\right\}.
\label{43}
\ees
\ni
This representation of the zero-frequency term
is in direct analogy with Eq.~(\ref{27}) for two plates. Like in (\ref{27}),
no discontinuity is contained in the single integral with respect to $y$.
As to the double integral in (\ref{43}), it is discontinuous with
respect to the relaxation parameter $\tilde\gamma$ at a point
$\tilde\gamma=0$.
In fact, it follows from Eq.~(\ref{40})
\beq
\frac{\partial f_{sl}^{(2)}(\tilde\xi,y)}{\partial{\tilde\xi}}=
\frac{2r_2(\tilde\xi,y)}{e^y-r_2^2(\tilde\xi,y)}
\frac{\partial r_2(\tilde\xi,y)}{\partial{\tilde\xi}},
\label{44}
\eeq
\ni
where the derivative of the transverse reflection coefficient was
calculated in Eq.~(\ref{30}). Once again, in the case of plasma
model the derivative (\ref{44}), and also the double integral in (\ref{43}),
vanish. As to the Drude model, it follows from (\ref{44})
\beq
\lim\limits_{\tilde\gamma\to 0}
\frac{\partial f_{sl}^{(2)}(\tilde\xi,y)}{\partial{\tilde\xi}}=0
\label{45}
\eeq
\ni
but
\beq
\lim\limits_{\tilde\gamma\to 0}
\int\limits_{0}^{y}d{\tilde\xi}
\frac{\partial f_{sl}^{(2)}(\tilde\xi,y)}{\partial{\tilde\xi}}=
\ln\left[1-\left(\frac{y-\sqrt{{\tilde\omega}_p^2+y^2}}{y+
\sqrt{{\tilde\omega}_p^2+y^2}}\right)^2e^{-y}\right]\neq 0,
\label{46}
\eeq
\ni
which means the discontinuity of the double integral in (\ref{43})
as a function of $\gamma$.

This discontinuity leads to the nonphysical results (see Secs.~IV,\,V).
It can be deleted by taking the limit
of the dielectric permittivity to infinity prior to setting the frequency
equal to zero in accordance with prescription used in
Sec.~II. Then due to (\ref{30}), (\ref{44})
\beq
\frac{\partial f_{sl}^{(2)}(\tilde\xi,y)}{\partial{\tilde\xi}}
\sim\frac{1}{\sqrt{\varepsilon-1}}{\to}0
\quad \mbox{with}{\ }{\varepsilon\to\infty}
\label{47}
\eeq
\ni
and the discontinuous term in (\ref{43}) disappears.

Using this prescription, one finally obtains the expression for
the Casimir force acting between a plate and a spherical lens (sphere)
at nonzero temperature
\bes
&&
F_{sl}(a)=\frac{k_BTR}{8 a^2}\left\{
\int\limits_{0}^{\infty}y\,dy
\left[f_{sl}^{(1)}(0,y)+f_{sl}^{(2)}(y,y)\right]\right.
\nn \\
&&\phantom{F_{ss}(a)}
\left.+2
\sum\limits_{l=1}^{\infty}
\int\limits_{{\tilde\xi}_l}^{\infty}y\,dy
f_{sl}({\tilde\xi}_l,y)\right\}.
\label{48}
\ees
\ni
The calculations of the Casimir force on the base of  Eq.~(\ref{48})
are contained in Sec.~V.
The obtained results are compared with those obtained
without the use of the above prescription \cite{29,30} or using
another prescription \cite{31,32}.

\section{COMPUTATIONS OF THE CASIMIR FORCE BETWEEN 
TWO PLATES MADE OF REAL METAL}

We start the numerical computation of the Casimir force with
Eq.~(\ref{35}) using the Drude and plasma
dielectric functions (\ref{12}).
For the sake of definiteness, we consider the case of aluminum
whose plasma frequency and relaxation parameter can be chosen
as follows \cite{45}
\bes
&&
\omega_p\approx 12.5\,\mbox{eV}\approx 1.9\times 10^{16}\,
\mbox{rad/s},
\label{49} \\
&&
\gamma\approx 0.063\,\mbox{eV}\approx 9.6\times 10^{13}\,
\mbox{rad/s}.
\nn
\ees
\ni
The most descriptive quantity illustrating the temperature
correction to the Casimir force is
\beq
\delta_T(F_{ss}^f)=
\frac{F_{ss}^f(a,T_0)-F_{ss}^f(a,0)}{F_{ss}^f(a,0)}.
\label{50}
\eeq
\ni
Here the upper index $f$ near the quantity $F_{ss}$ from
Eq.~(\ref{35}) signifies one or another
dielectric function (Drude or plasma for instance),
and the second argument $T_0=300\,$K, and $T=0$ means
the temperature at which the Casimir force is computed.
The relative temperature correction (\ref{50}) is a function
of separation $a$. It incorporates effectively both the case
of low temperatures (which occurs when $T_0\ll T_{eff}$ at small
$a$) and high temperatures (as $T_0\gg T_{eff}$ at large $a$).

\begin{figure}[h]
\epsfxsize=15cm\centerline{\epsffile{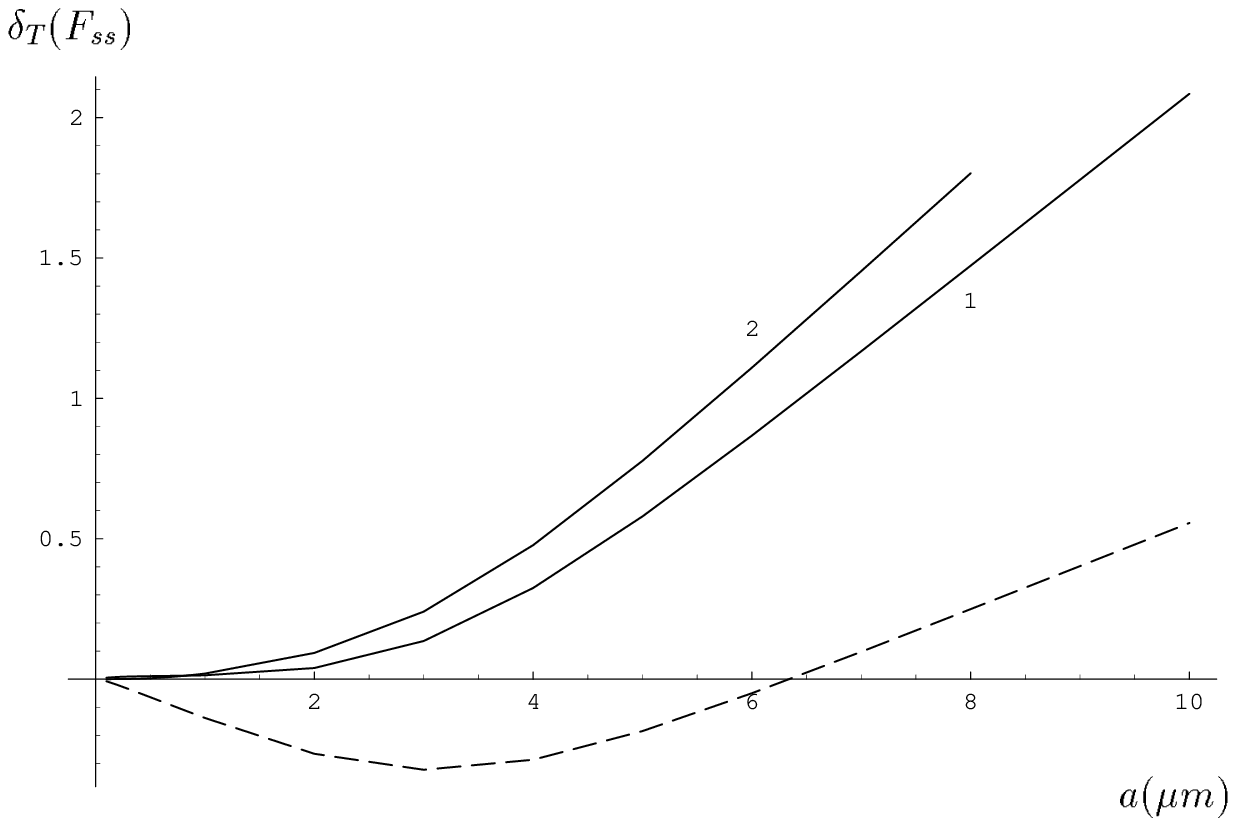}}
\vspace*{-4cm}
\caption{\label{forceT}
Relative temperature correction to the Casimir
force between two plates in dependence of separation.
Curve 1 corresponds to Drude model (our computation),
the dashed curve is obtained in Drude model with
$r_2(0,k_{\perp})=0$, and curve 2 is for the dielectric
plates.
}
\end{figure}

The results of numerical computations by Eqs.~(\ref{35}),
(\ref{50}) in the case $f=D$ (Drude mo\-del)
are presented in Fig.~\ref{forceT} by the solid
curve 1. In the same figure the solid curve 2 illustrates the
result obtained for dielectric plates (this case is discussed
in Sec.~VI). The dashed curve represents the result 
obtained
by the approach of \cite{29,30}. In accordance with this approach,
as was discussed in Sec.~II, one should substitute
$f_{ss}^{(2)}(0,y)=0$ instead of $f_{ss}^{(2)}(y,y)$ into
Eq.~(\ref{35}).

As is seen from Fig.~\ref{forceT} (curve 1), in the case
of real metals the relative temperature correction is
rather small at small separations and monotonically
increases with the increase of separation remaining
positive as it should be from general thermodynamical
considerations. In contrast to this, the temperature
correction, computed on the basis of \cite{29,30}, is
negative in a wide range of separations and turns into zero
at a separation $a\approx 6.3\,\mu$m. For larger distances
it becomes positive. At small distances it is rather
large by the modulus and increases linearly with distance
(temperature). This behavior is in radical contradiction
with both the case of ideal metal and a metal described
by the plasma model dielectric function considered in detail
in \cite{27,28}. At the same time the results obtained here
are in good agreement with both limiting cases of
an ideal metal and metal described by the plasma model.

Note that in the original paper \cite{30} not $\delta_T(F)$
but the correction factor was plotted as a function of
separation $a$, i.e. the ratio between the Casimir force
for real metals computed at $T_0=300\,$K and the Casimir
force for ideal metals at zero temperature. According to
the results of \cite{27} in the case of plasma model
this correction factor is approximately equal to the
product of the correction factors through finite conductivity
(at zero temperature) and through nonzero temperature
(for the ideal metal). Thus the resulting negative
temperature corrections, which are unacceptable from the
thermodynamical point of view, were not clearly evident in \cite{30}.

In Table 1 the values of the relative temperature 
corrections are given at several distances between 
the plates
in the case of Drude model (present paper, column 2),
plasma model (present paper, column 3), for the ideal
metal (column 4), computed on the base of \cite{30} in
frames of Drude model (column 5), and computed on the
base of \cite{31,32} in frames of Drude and plasma
models (columns 6 and 7 respectively). We remind that in
\cite{31,32} all computations were performed 
by Eq.~(\ref{5}) where in the zeroth term the reflection coefficients
(\ref{16}) were substituted.
As is seen from the comparison of columns 2--4, in
the framework of the Drude model the relative temperature correction
at smallest separations is three orders larger than 
with the plasma model and four orders larger 
than for the
ideal metal. Note that for the ideal metal at small
separations the lowest nonzero temperature
correction to the zero-temperature result is of order
$(T/T_{eff})^4$ (see below). 
With increasing of separation
the difference between the predictions of Drude and plasma
models decreases approaching the values obtained for the
ideal metal (column 4). At small separations 
the modulus of temperature correction
calculated on the base of \cite{30} (column 5) is several
times larger than calculated by us (column 2) and 
the correction itself is negative.
The results of the calculations on
the base of \cite{31,32} (columns 6,\,7) are also
significantly different from our corresponding results
of columns 2,\,3. According to \cite{31,32} there 
exist large temperature corrections at small separations
which are linear in temperature. 

\begin{figure}[h]
\epsfxsize=15cm\centerline{\epsffile{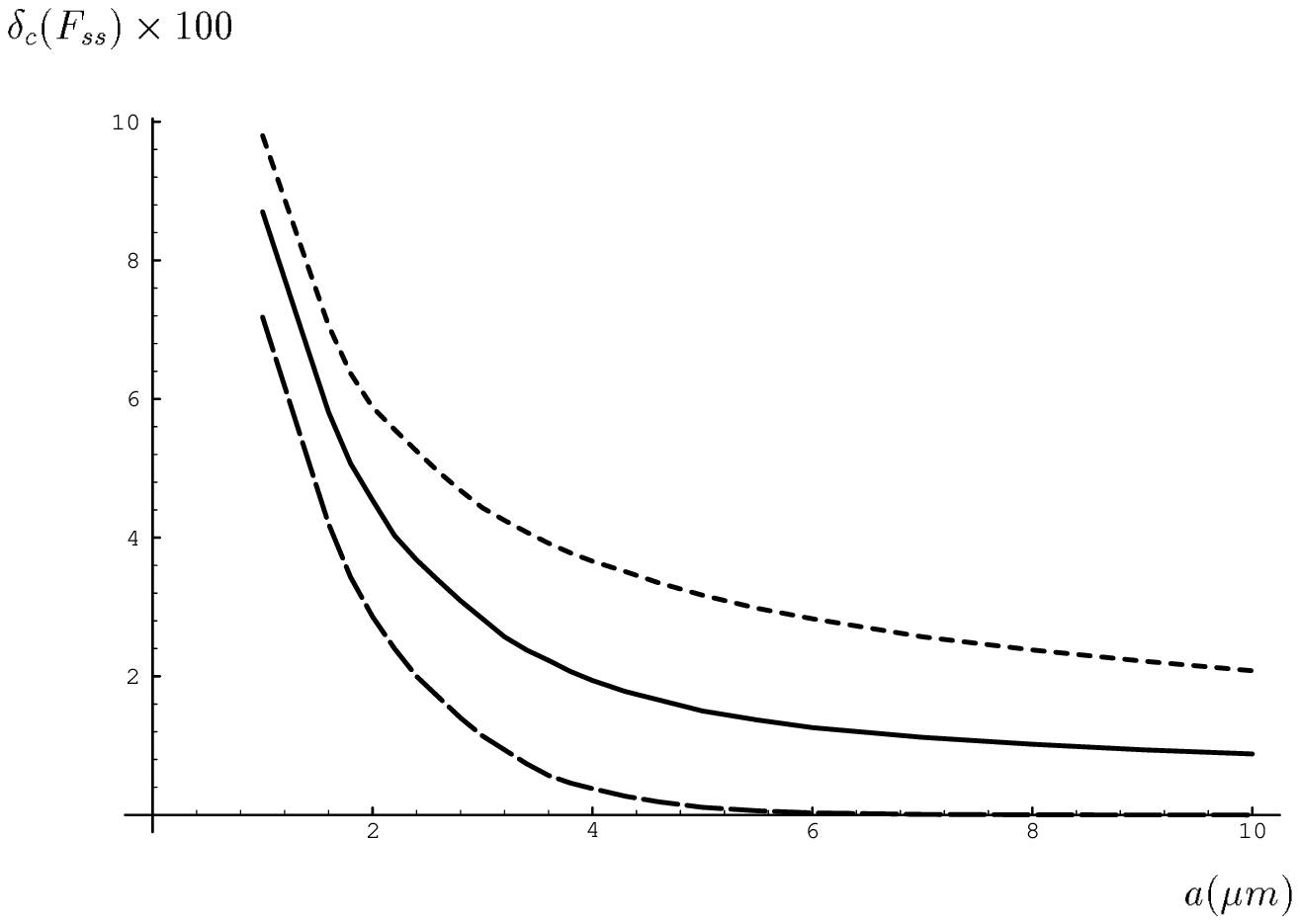}}
\vspace*{-4cm}
\caption{\label{force-dr}
Relative finite conductivity correction to the 
Casimir force between two plates
in dependence of separation in Drude model.
Solid curve represents our computations at $T=300\,$K,
long-dashed curve is obtained under the supposition
$r_{1,2}(0,k_{\perp})=1$ at $T=300\,$K,
short-dashed curve is for $T=0$.
}
\end{figure}

\begin{figure}[h]\epsfxsize=15cm\centerline{\epsffile{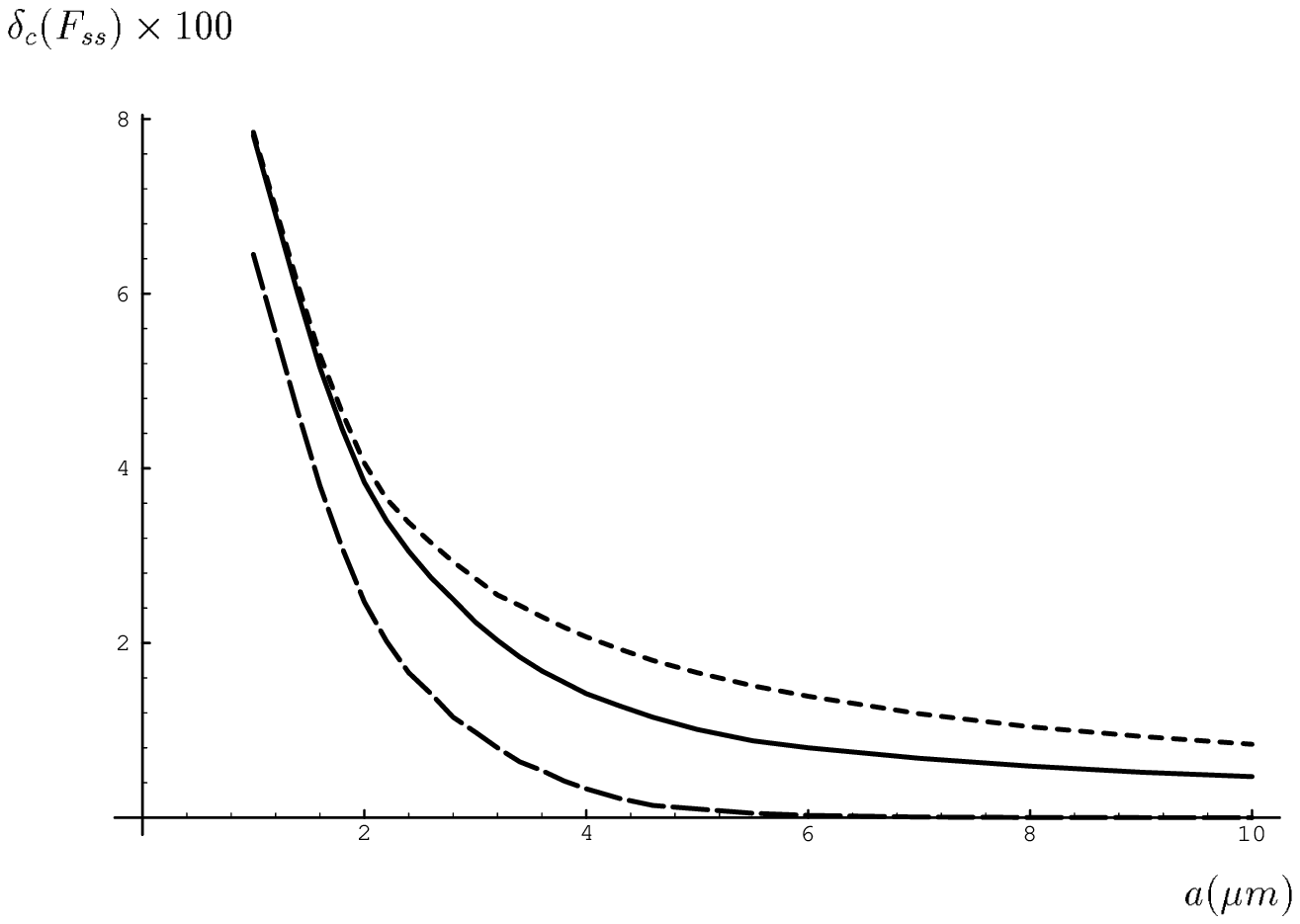}}
\vspace*{-4cm}
\caption{\label{force-pl}
Relative finite conductivity correction to the 
Casimir force between two plates
in dependence of separation in plasma model.
Solid curve represents our computations at $T=300\,$K,
long-dashed curve is obtained under the supposition
$r_{1,2}(0,k_{\perp})=1$ at $T=300\,$K,
short-dashed curve is for $T=0$.
}
\end{figure}

Now let us discuss the finite conductivity corrections
to the Casimir force at nonzero temperature. They can be
characterized by the quantity
\beq
\delta_c(F_{ss}^f)=
\frac{F_{ss}^0(a,T_0)-F_{ss}^f(a,T_0)}{F_{ss}^0(a,T_0)},
\label{51}
\eeq
\ni
where, as above, $F_{ss}^0(a,T_0)$ is the Casimir force
between perfect metallic plates at temperature $T_0$.
The results of numerical computations of 
$\delta_c(F_{ss}^f)$ in the case of the Drude model ($f=D$)
are shown in Fig.~\ref{force-dr}. The solid curve represents
the finite conductivity correction obtained by our
method (Eq.~(\ref{35})) at $T_0=300\,$K. For comparison
the corresponding results at zero temperature $T_0=0$
are given by the short-dashed curve. The long-dashed
curve is obtained at $T_0=300\,$K in the framework of the
approach of \cite{31,32}. In Fig.~\ref{force-pl} the
results computed in the case of the plasma model ($f=p$) are
shown with the same notations as in 
Fig.~\ref{force-dr}. As is seen from
Figs.~\ref{force-dr},\,\ref{force-pl}, in the Drude model the
finite conductivity correction is larger than in the  
plasma model at all temperatures and at all separations.
From Fig.~\ref{force-pl} it follows that at $a=1\,\mu$m
the finite conductivity correction computed using the
plasma model at $T_0=300\,$K is practically the same
as at $T_0=0$.

It is notable that the finite conductivity correction
computed on the basis of \cite{31,32} turns into zero
at separations larger than $5\,\mu$m in both Drude and
plasma models (see the long-dashed curves in
Figs.~\ref{force-dr} and \ref{force-pl}). In fact, as is seen
from the solid and short-dashed curves in
Figs.~\ref{force-dr},\,\ref{force-pl}, with increasing
temperature the finite conductivity correction
decreases. The extent of this decrease depends on the
model used ($D$ or $p$) and on the values of plasma and
relaxation frequencies. In contrast to this, the finite
conductivity correction computed on the basis of
\cite{31,32} turns into zero at one and the same separation
not only in different models but also for one and the same
model with different values of parameters, i.e.
regardless of the quality of metal under consideration.
It is easily seen, that the separation value, at which
the finite conductivity corrections of \cite{31,32}
vanish, is determined not by the conductivity properties
of a metal but by the smallness of terms with
$l\geq 1$ in the Lifshitz formula which are of order
$\exp(-2\pi T/T_{eff})$. This means that the anzats
used in \cite{31,32} to redefine the zeroth term of
Lifshitz formula (see the above discussion in Sec.~II)
is unjustified.

\begin{figure}[h]
\epsfxsize=15cm\centerline{\epsffile{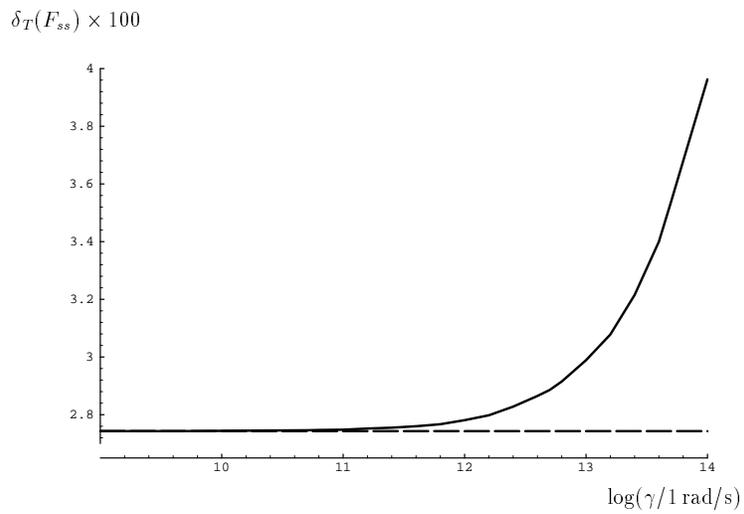}}
\vspace*{-4cm}
\caption{\label{dr-pl}
Relative temperature correction to the Casimir force
between two plates in dependence of relaxation
frequency at a separation $a=2\,\mu$m and $T=300\,$K
(solid curve). The dashed line is for plasma model.
}
\end{figure}

It is of interest to consider the dependence of our
results on the relaxation parameter $\gamma$ of Drude
model.
In Fig.~\ref{dr-pl} the relative temperature correction
computed by Eqs.~(\ref{35}), (\ref{50}) is plotted with
the decreasing of $\gamma$ at a separation $a=2\,\mu$m,
and at a temperature $T_0=300\,$K (solid curve).
In the same figure the dashed curve represents the
result obtained using the plasma model. It is clearly
seen that with the decrease of $\gamma$ by 3 orders of
magnitude the results computed in the Drude model join
smoothly with the results of the plasma model as it should
be from general considerations. Note that this is not
the case in the approach of \cite{29,30} where the 
results obtained with the Drude model do not join
the results of the plasma model with decreasing $\gamma$.
The reason is that, different definitions of the
zero-frequency term of Lifshitz formula are used in
\cite{29,30} and in the present paper (see Sec.~II).

As to the papers \cite{31,32}, there is the smooth
transition between the results obtained in Drude and
plasma models. However, as was indicated in
Figs.~\ref{force-dr},\,\ref{force-pl}, for 
$a>4\,\mu$m the results in both models coincide with one
another and for separations $a>6\,\mu$m coincide also
with the case of ideal metal.

Now let us derive some analytic results for the configuration
of two plane parallel plates using both the plasma
and Drude models. In the plasma model it is possible to
obtain the perturbation expansion of Eq.~(\ref{23}) 
in terms of a small parameter $\delta_0/a$, where
$\delta_0=c/\omega_p$ is the effective penetration depth
of the electromagnetic zero-point oscillations into the metal.
For this purpose it is useful to introduce the new
variable $v=\tilde\xi/y$ instead of $\tilde\xi$ and
to rewrite Eq.~(\ref{23}) in the form
\beq
F_{ss}(a)=
-\frac{\hbar c}{16\pi^2 a^4}
\sum\limits_{l=0}^{\infty}{\vphantom{\sum}}^{\prime}
\int\limits_{0}^{\infty}y^3\,dy
\int\limits_{0}^{1}dv
\cos\left(lvyt\right)
f_{ss}(v,y).
\label{52}
\eeq
\ni
Expanding the quantity $f_{ss}$ defined in (\ref{20})
up to the first order in powers of $\delta_0/a$ one
obtains
\beq
f_{ss}(v,y)=\frac{2}{e^y-1}-2\frac{ye^y}{(e^y-1)^2}
\left(1+v^2\right)\frac{\delta_0}{a}.
\label{53}
\eeq
\ni
Substituting Eq.~(\ref{53}) into Eq.~(\ref{52}) we come
after some transformations to the Casimir force including
the effect of both the nonzero temperature and finite
conductivity
\bes
&&
F_{ss}(a)=F_{ss}^0(a)\left\{1+
\frac{30}{\pi^4}
\sum\limits_{l=1}^{\infty}
\left[\frac{1}{t^4l^4}-\frac{\pi^3}{tl}
\frac{\cosh(\pi tl)}{{\sinh}^3(\pi tl)}\right]
\right.
\label{54}\\
&&\phantom{aaaa}
-\frac{16}{3}\frac{\delta_0}{a}-
60\frac{\delta_0}{a}
\sum\limits_{l=1}^{\infty}
\left[\frac{2{\cosh}^2(\pi tl)+1}{{\sinh}^4(\pi tl)}
-\frac{2{\cosh}(\pi tl)}{{\pi tl\sinh}^3(\pi tl)}
\right.
\nn \\
&&\phantom{aaaa}
\left.\left.
-\frac{1}{2\pi^2t^2l^2{\sinh}^2(\pi tl)}
-\frac{\coth(\pi tl)}{2\pi^3t^3l^3}\right]\right\},
\nn
\ees
\ni
where 
$F_{ss}^0(a)= F_{ss}^0(a,0)\equiv -\pi^2\hbar c/(240a^4)$
is the zero-temperature Casimir force between ideal
metals. The first summation in (\ref{54}) is exactly the
temperature correction in the case of ideal metals
(see, e.g., \cite{17}). The second summation takes into
account the effect of finite conductivity combined with
nonzero temperature.

In the limit of low temperatures $T\ll T_{eff}$ one has
from (\ref{54}) neglecting terms exponentially small in
$2\pi T_{eff}/T$ \cite{28}
\bes
&&
F_{ss}(a)\approx F_{ss}^0(a)\left\{1+
\frac{1}{3}\left(\frac{T}{T_{eff}}\right)^4
\right.
\label{55} \\
&&\phantom{aaaa}\left.
-\frac{16}{3}\frac{\delta_0}{a}
\left[1-\frac{45\zeta(3)}{8\pi^3}
\left(\frac{T}{T_{eff}}\right)^3\right]\right\}.
\nn
\ees
\ni
For $\delta_0=0$ (perfect conductor) Eq.~(\ref{55}) turns
into the well known result \cite{17} demonstrating
that the first nonzero temperature correction is of
the fourth power in $T/T_{eff}$. For $T=0$ the 
first-order finite conductivity correction 
to the Casimir
force \cite{16,17} is reproduced from (\ref{55}).
Note that the first correction of mixing finite
conductivity and finite temperature is of order
$(T/T_{eff})^3$. More significantly, note  that there are
no temperature corrections of order $(T/T_{eff})^k$
with $k\leq 4$
in the higher order conductivity correction terms
$(\delta_0/a)^i$ from the second up to the sixth
order \cite{28}.

In the Drude model the analytical results can be obtained
in the high-temperature limit $T\gg T_{eff}$. It is easily
seen that at high temperatures (large separations) only
one term of Eq.~(\ref{35}) with $l=0$ contributes the
result, the other terms with $l\geq 1$ being exponentially
small in the parameter $2\pi T/T_{eff}$. The situation
here is exactly the same as for the ideal metal \cite{17}.
As a consequence, Eq.~(\ref{35}) can be rewritten in the
following form
\beq
F_{ss}(a)=-\frac{k_BT}{16\pi a^3}
\left[
\int\limits_{0}^{\infty}
\frac{y^2\,dy}{e^y-1}+
\int\limits_{0}^{\infty}
\frac{y^2\,dy}{r_2^{-2}(y,y)e^y-1}\right].
\label{56}
\eeq
\ni
It is seen from Eq.~(\ref{56}) that in the high-temperature
limit only the perpendicular reflection coefficient
$r_2$ gives rise to the finite conductivity corrections
to the Casimir force (the same is valid in plasma model
also).

After the straightforward calculations up to the first
orders in small parameters $\delta_0/a$,
$\gamma/\omega_p$ (see the details in Appendix II) the
result  
\beq
F_{ss}(a)=F_{ss}^0(a,T)\left[1-3\frac{\delta_0}{a}-
\frac{\gamma}{\omega_p}\frac{1}{\zeta(3)}
I_1(\tilde\gamma)\right]
\label{57}
\eeq
\ni
is obtained,
where $I_1$ is a function slowly depending on the effective
relaxation parameter (space separation) and given by
\beq
I_1(\tilde\gamma)=
\int\limits_{0}^{\infty} dy\,
\frac{y^2\sqrt{y}}{\sqrt{y+\tilde\gamma}+\sqrt{y}}\,
\frac{e^y}{(e^y-1)^2}.
\label{58}
\eeq
\ni
At high temperatures, which are considered here, the
Casimir force acting between the ideal metals is
\beq
F_{ss}^0(a,T)=-\frac{k_BT}{4\pi a^3}\zeta(3),
\label{59}
\eeq
\ni
where $\zeta(3)\approx 1.202$ is the Riemann zeta
function.

With the above value of relaxation frequency for $Al$
one has $I_1\approx 1.16$ at $a=5\,\mu$m and
$I_1\approx 0.99$ at $a=10\,\mu$m. In Appendix II
(Fig.~\ref{dr-ht})
the functional dependence of $I_1$ on $\tilde\gamma$
is plotted. As is seen from (\ref{57}), the
high-temperature Casimir force depends on both plasma
frequency and relaxation frequency. For all
$a>6\,\mu$m the results obtained by the asymptotic
Eq.~(\ref{57}) coincide with the above results of
numerical computations (see Figs.~\ref{force-dr} 
and \ref{force-pl}). 
The characteristic size of the conductivity correction
at large separations can be estimated from the following
example. At $a=10\,\mu$m
the finite conductivity correction obtained from (\ref{57})
is $\delta_c(F_{ss}^D)\approx 0.89$\% and with $\gamma=0$
it holds
$\delta_c(F_{ss}^p)\approx 0.47$\% in perfect agreement with
Figs.~\ref{force-dr},\,\ref{force-pl}. The smooth
joining of the results obtained using the Drude model
with those using the plasma model when $\gamma\to 0$ is evident.

Note that in the framework of the approach used in \cite{29,30}
the high temperature Casimir force between real metals
is given by $F_{ss}(a)=F_{ss}^0(a,T)/2$ (see 
Eq.~(\ref{59})), i.e. two times smaller than for 
ideal metal, regardless of the conductivity properties
of a real metal used. As to the approach used in \cite{31,32},
the asymptotic behavior at high temperatures coincides with
Eq.~(\ref{59}), i.e. it is the same as the case for ideal metals in
both plasma and Drude models. Once again, the real 
properties of a metal do not influence the result.

\section{COMPUTATIONS OF THE CASIMIR FORCE BETWEEN 
A PLATE AND A SPHERICAL LENS MADE OF REAL METAL}

The configuration of a spherical lens (or a sphere) above
a semispace (plate) was found to be the most suitable for
the precision measurements of the Casimir force
\cite{7,8,9,10,11,11a}. In these experiments the finite
conductivity corrections have been demonstrated 
\cite{8,9,10,11,11a} and the sensitivity for the detection of the 
temperature corrections is close to being achieved \cite{7}. For this
reason the
combined effect of both corrections is of extreme
interest. Here the computations of the Casimir force
for the  configuration of a lens above a plate are performed
using the Drude and plasma models with parameters of
Eq.~(\ref{49}). The results obtained by our Eq.~(\ref{48})
are compared with the computations of other authors.

\begin{figure}[h]
\epsfxsize=15cm\centerline{\epsffile{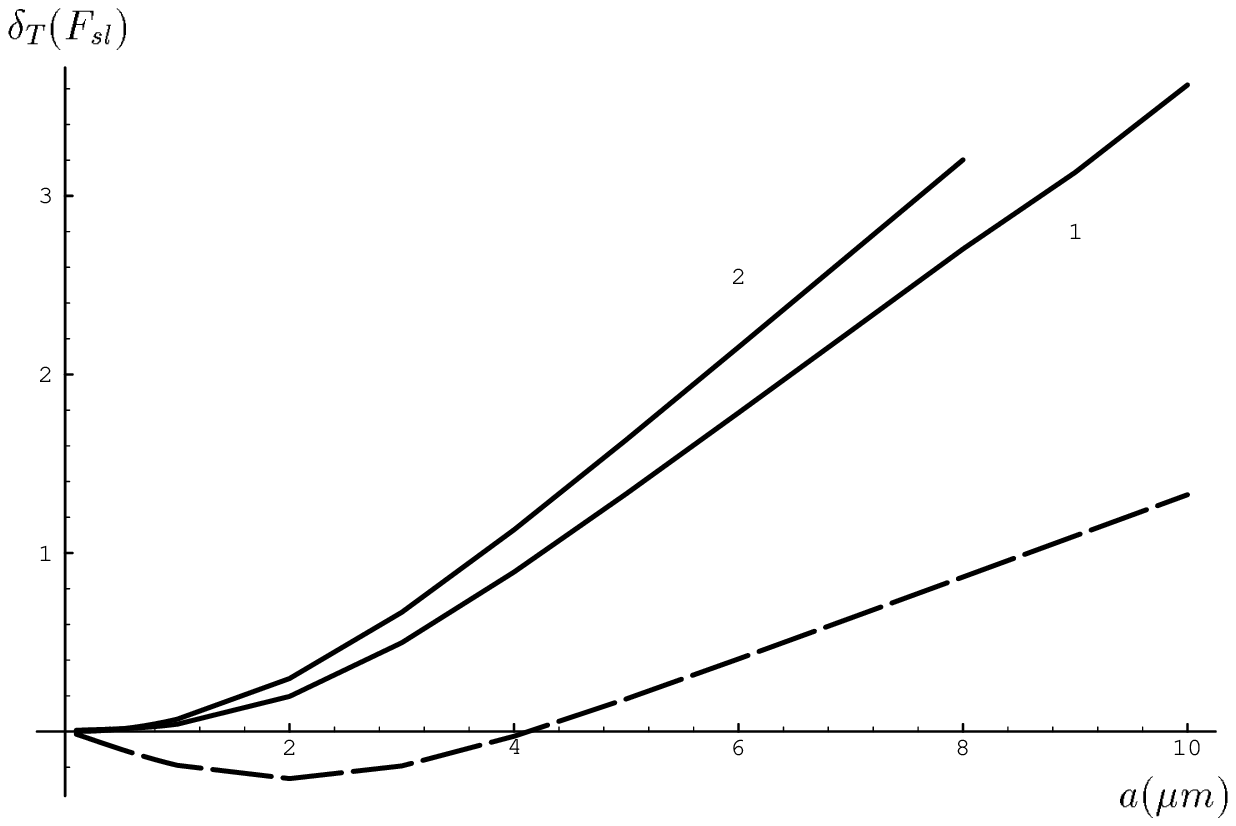}}
\vspace*{-4cm}
\caption{\label{energyT}
Relative temperature correction to the Casimir
force between a plate and a lens in dependence of 
separation.
Curve 1 corresponds to Drude model (our computation),
the dashed curve is obtained in Drude model with
$r_2(0,k_{\perp})=0$, and curve 2 is for the dielectric
test bodies.
}
\end{figure}

We start with the relative temperature correction
\beq
\delta_T(F_{sl}^f)=
\frac{F_{sl}^f(a,T_0)-F_{sl}^f(a,0)}{F_{sl}^f(a,0)},
\label{60}
\eeq
\ni
where all the notations and parameters are the same as
in Eq.~(\ref{50}) and only the configuration is different.
The results of numerical computations by Eqs.~(\ref{48}),
(\ref{60}) in the case of Drude model are presented in 
Fig.~\ref{energyT} by the solid curve 1. The solid curve 2
represents the temperature correction in the case of
a dielectric plate and a lens (see the next section).
By the dashed curve the results obtained by the approach
of \cite{29,30} are shown. The curve 1 increases
monotonically in perfect analogy with Fig.~\ref{forceT}.
However, the dashed curve represents the negative
temperature correction at separations $a\leq 4.1\,\mu$m
and changes sign for larger separations. This behavior
corresponds to the case where large by modulus 
corrections linear in temperature to the ideal 
zero-temperature Casimir force are present. According to \cite{33}
such corrections are in contradiction with the
experimental data of \cite{7}. On the basis of our
Eq.~(\ref{48}) such corrections do not arise.

In Table 2 the values of the relative temperature
corrections are presented at different distances
between a lens and a plate computed using the
Drude model (present paper, column 2) and the plasma model
(present paper, column 3),  the ideal metal (column 4),
computed on the basis of \cite{30} using the Drude
model (column 5), and on the basis of \cite{31,32} using 
Drude and plasma models (columns 6 and 7 respectively).

From Table 2 (columns 2--4) it follows that at smallest
separations the temperature correction computed using the
Drude model is about two orders of magnitude larger than with 
the plasma model and for the ideal metal. At large
separations the predictions of both models are very close 
to each other and to the results obtained with the ideal metal 
(column 4). The negative and large magnitude of the corrections at small
separations results of column 5, computed on the basis
of \cite{30}, correspond to linear temperature
corrections. The results of columns 6,\,7, although
positive, also correspond to the presence of linear
temperature corrections at small separations.

\begin{figure}[h]
\epsfxsize=15cm\centerline{\epsffile{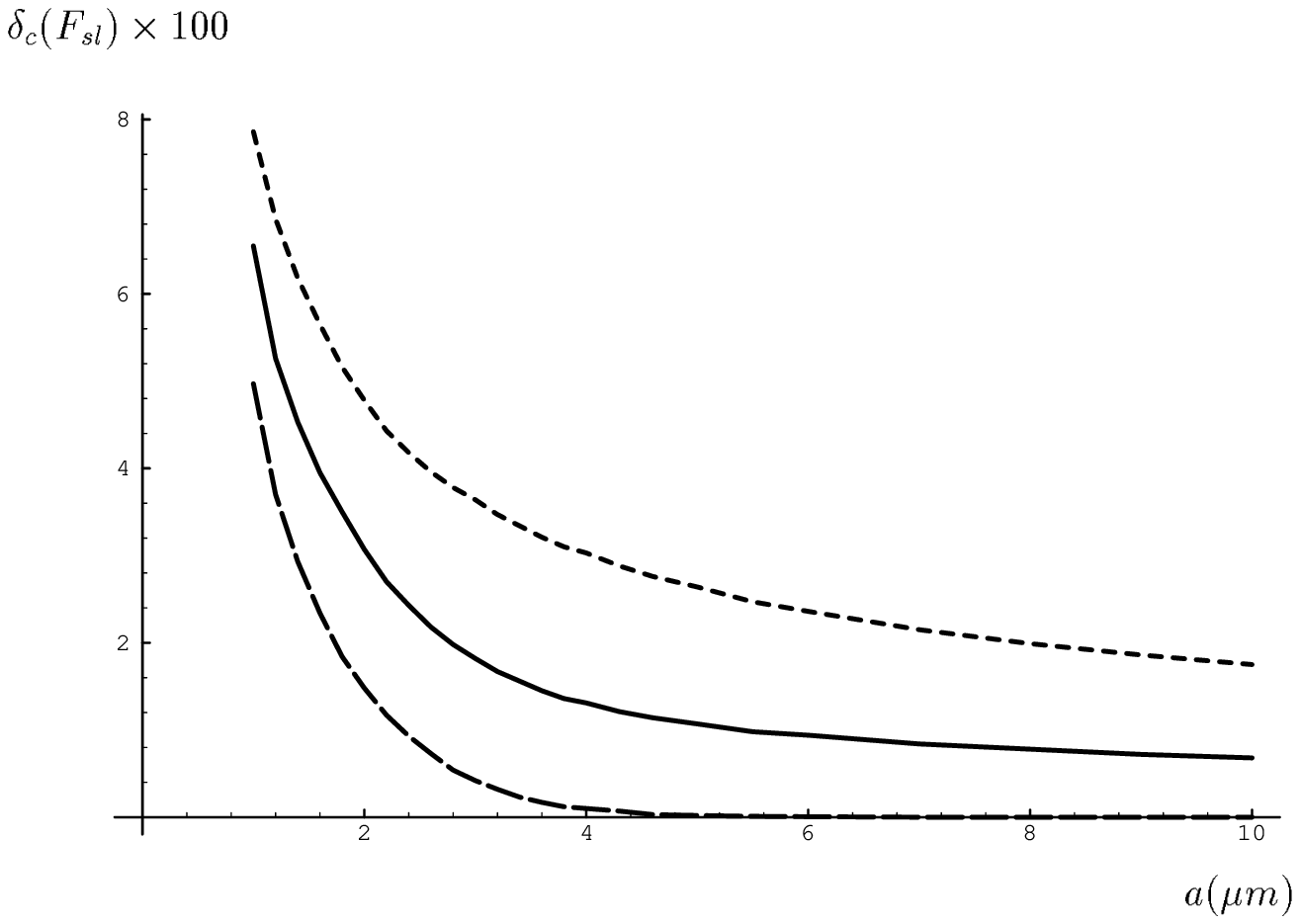}}
\vspace*{-4cm}
\caption{\label{energ-dr}
Relative finite conductivity correction to the 
Casimir force between a plate and a lens
in dependence of separation in Drude model.
Solid curve represents our computations at $T=300\,$K,
long-dashed curve is obtained under the supposition
$r_{1,2}(0,k_{\perp})=1$ at $T=300\,$K,
short-dashed curve is for $T=0$.
}
\end{figure}

\begin{figure}[h]
\epsfxsize=15cm\centerline{\epsffile{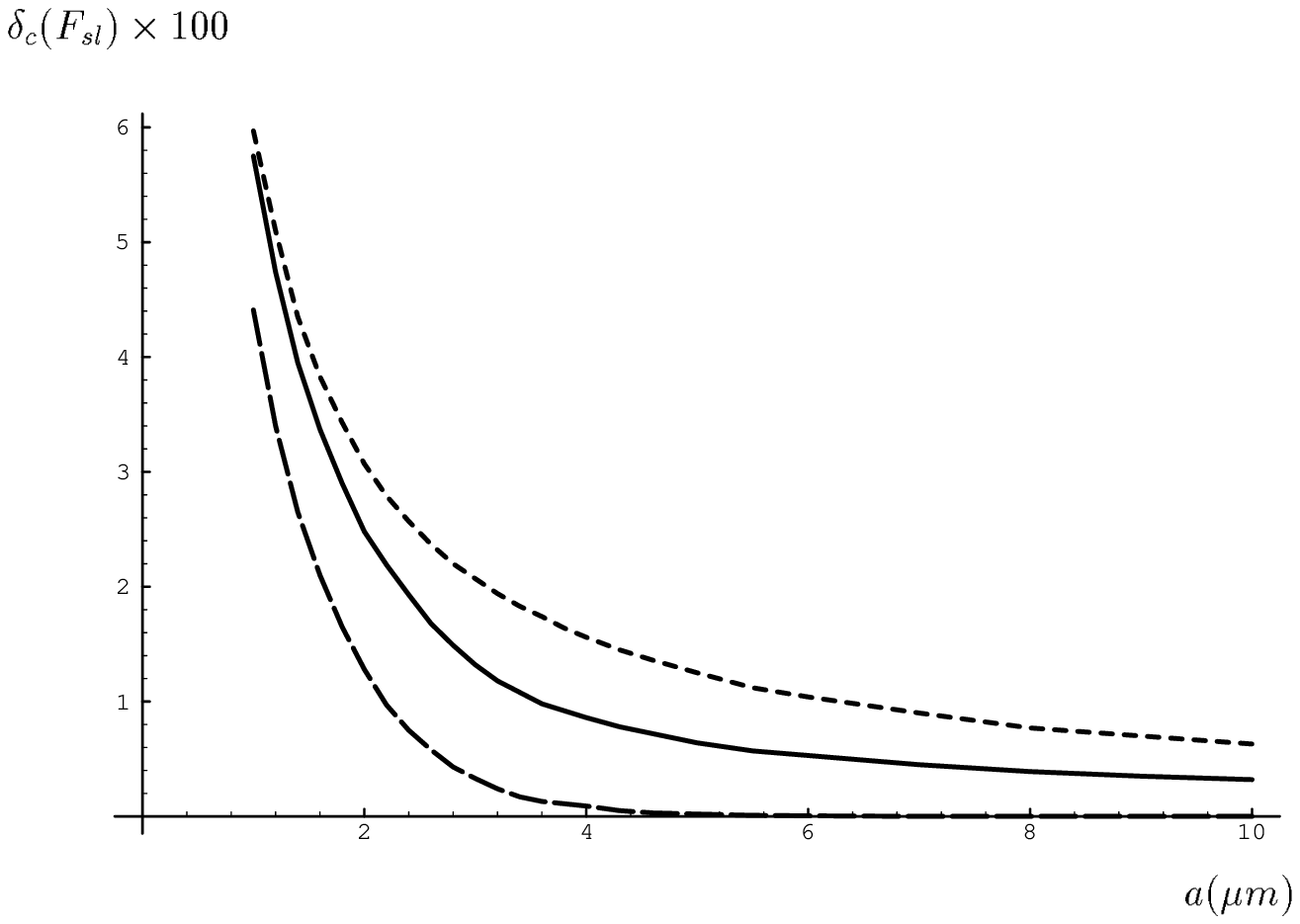}}
\vspace*{-4cm}
\caption{\label{energ-pl}
Relative finite conductivity correction to the 
Casimir force between a plate and a lens
in dependence of separation in plasma model.
Solid curve represents our computations at $T=300\,$K,
long-dashed curve is obtained under the supposition
$r_{1,2}(0,k_{\perp})=1$ at $T=300\,$K,
short-dashed curve is for $T=0$.
}
\end{figure}

The relative finite conductivity correction at a
temperature $T_0$ can be described by
\beq
\delta_c(F_{sl}^f)=
\frac{F_{sl}^0(a,T_0)-F_{sl}^f(a,T_0)}{F_{sl}^0(a,T_0)}.
\label{61}
\eeq
\ni
In Fig.~\ref{energ-dr} this quantity is plotted as a
function of separation in the case of the Drude model at
$T_0=300\,$K (solid curve is our result 
computed by Eq.~(\ref{48}),
the long-dashed curve is computed on the basis of
\cite{31,32}). The dependence of the conductivity
correction on separation at $T_0=0$ is shown by the
short-dashed curve. In Fig.~\ref{energ-pl} the analogous
results computed using the plasma model are presented.
As with the case of two plane plates (Sec.~IV),
Drude model leads to larger finite conductivity corrections
than the plasma model. At separations
larger than $4\,\mu$m the conductivity correction
computed using \cite{31,32} turns into zero with
both Drude and plasma models (the long-dashed curves in
Figs.~\ref{energ-dr}, \ref{energ-pl}). Once more, this
property is determined by the artificial modification
of the zeroth term of Lifshitz formula used 
in \cite{31,32} and does not depend on the particular
characteristics of a real metal.

In the same way as for two plane plates our results for
a lens above a plate, computed using the Drude model,
join smoothly when $\gamma\to 0$ with the results
computed with the plasma model. This is not the case in the
approach of \cite{29,30}.

We come now to the perturbative analytical results
which can be obtained for the configuration of a lens
above a plate. Here, the plasma model can be
used. Introducing the new variable 
$v=\tilde\xi/y$ instead of $\tilde\xi$ one obtains
\beq
F_{sl}(a)=\frac{\hbar cR}{8\pi a^3}
\sum\limits_{l=0}^{\infty}{\vphantom{\sum}}^{\prime}
\int\limits_{0}^{\infty}y^2\,dy
\int\limits_{0}^{1}dv\,\cos(ltvy)f_{sl}(v,y).
\label{62}
\eeq
\ni
The expansion of $f_{sl}$ up to first order in the small
parameter $\delta_0/a$ is
\beq
f_{sl}(v,y)=2\ln\left(1-e^{-y}\right)+
2\frac{y}{e^y-1}(1+v^2)\frac{\delta_0}{a}.
\label{63}
\eeq
\ni
Substitution of Eq.~(\ref{63}) into Eq.~(\ref{60})
leads to result
\bes 
&&F_{sl}(a)=F_{sl}^0(a)\left\{1+
\frac{45}{\pi^3}
\sum\limits_{l=1}^{\infty}
\left[\frac{\coth(\pi tl)}{t^3l^3}+
\frac{\pi}{t^2l^2{\sinh}^2(\pi tl)}\right]
-\frac{1}{t^4}\right.
\nn \\
&&\phantom{a}
\left.
-4\frac{\delta_0}{a}+\frac{180}{\pi^4}\frac{\delta_0}{a}
\sum\limits_{l=1}^{\infty}
\left[\frac{\pi\coth(\pi tl)}{2t^3l^3}
-\frac{2}{t^4l^4}+
\frac{\pi^3\coth(\pi tl)}{tl{\sinh}^2(\pi tl)}\right.\right.
\label{64} \\
&&\phantom{aaaaaaaaaaaaaaaaaa}
\left.\left.
+\frac{\pi^2}{t^2 l^2{\sinh}^2(\pi tl)}
\right]\right\},
\nn
\ees
\ni
where 
$F_{sl}^0(a)=F_{sl}^0(a,0)\equiv -\pi^3\hbar cR/(360a^3)$.

For the case of low temperatures $T\ll T_{eff}$ 
Eq.~(\ref{64}) leads to \cite{28}
\bes
&&
F_{sl}(a)\approx F_{sl}^0(a)\left\{1+
\frac{45\zeta(3)}{\pi^3}\left(\frac{T}{T_{eff}}\right)^3
-\left(\frac{T}{T_{eff}}\right)^4\right.
\nn \\
&&\phantom{aa}
\left.
-4\frac{\delta_0}{a}\left[1-
\frac{45\zeta(3)}{2\pi^3}\left(\frac{T}{T_{eff}}\right)^3
+\left(\frac{T}{T_{eff}}\right)^4\right]\right\}.
\label{65}
\ees
\ni
The corrections to (\ref{65}) are exponentially small
in the parameter $2\pi T_{eff}/T$.
For ideal metal $\delta_0=0$ and Eq.~(\ref{65}) coincides
with the known result \cite{7}. It is seen that for the 
perfectly conducting lens and plate the first nonzero
temperature correction is of the third order in
$T/T_{eff}$. For $T=0$ the first order finite conductivity 
correction to the Casimir force \cite{19}
is reproduced. In analogy with two plane plates the
perturbation orders $(\delta_0/a)^i$ with 
$2\leq i\leq 6$ do not contain temperature corrections of
orders $(T/T_{eff})^3$ and $(T/T_{eff})^4$ 
or smaller ones \cite{28}.

Now we consider the analytical results which can be
obtained with the Drude model at high temperature $T\gg T_{eff}$.
Here only the zeroth term of Lifshitz formula contributes
the result. Then Eq.~(\ref{48}) can be represented as
\beq
F_{sl}(a)=\frac{k_BTR}{8 a^2}
\left\{
\int\limits_{0}^{\infty}
y\,dy\,\ln(1-e^{-y})+
\int\limits_{0}^{\infty}
y\,dy\,\ln\left[1-r_2^{2}(y,y)e^{-y}\right]\right\}.
\label{66}
\eeq

After some transformations (see Appendix II) one arrives
at the result
\beq
F_{sl}(a)=F_{sl}^0(a,T)\left[1-2\frac{\delta_0}{a}-
\frac{\gamma}{\omega_p}\frac{2}{\zeta(3)}
I_2(\tilde\gamma)\right],
\label{67}
\eeq
\ni
where $I_2$ is defined by
\beq
I_2(\tilde\gamma)=
\int\limits_{0}^{\infty} dy\,
\frac{y\sqrt{y}}{\sqrt{y+\tilde\gamma}+\sqrt{y}}\,
\frac{1}{e^y-1}.
\label{68}
\eeq
\ni
The high-temperature Casimir force acting between the 
lens and the plate made of ideal metal is
\beq
F_{sl}^0(a,T)= -\frac{k_BTR}{4a^2}\zeta(3).
\label{69}
\eeq

For example,
$I_2\approx 0.519$ at $a=5\,\mu$m and
$I_1\approx 0.434$ at $a=10\,\mu$m
(using the data of (\ref{49}) for $Al$). 
The dependence of $I_2$ on $\tilde\gamma$ is plotted
in Appendix II (Fig.~\ref{dr-ht}). The asymptotic results
of Eq.~(\ref{67}) coincide with the results of numerical
computations for $a>5\,\mu$m. The choice of the model
describing dielectric properties of a metal at large
separations is rather important.
At $a=10\,\mu$m the conductivity correction 
is $\delta_c(F_{sl}^D)\approx 0.68$\% using the Drude
model and $\delta_c(F_{sl}^p)\approx 0.32$\% using
plasma model, i.e. more than two times smaller.
In analogy with two plates, the high-temperature Casimir
force calculated on the basis of \cite{30}
is two times smaller than in (\ref{69}) regardless of 
the conductivity properties of the metal. In \cite{31,32}
the high-temperature behavior for the real metals
coincides with (\ref{69}) obtained for the ideal metal.
So, the actual properties of a particular metal are
not reflected.

\section{TEMPERATURE CASIMIR FORCE BETWEEN
THE DIELECTRIC TEST BODIES}

Here we briefly discuss the temperature Casimir force
between dielectrics. The
application of Lifshitz formulas (\ref{1}) or (\ref{37})
for the case of dielectric surfaces is direct. No additional
prescriptions, such as the ones used above 
or their generalizations
are needed to obtain the final result matching the
general physical requirements. However, when the formalism
developed for dielectrics is compared with that for metals,
the origin of the above difficulties becomes clear.

The dielectric permittivity of dielectrics can be
modeled, e.g., by the Mahanty-Ninham relation
\cite{41,49}
\beq
\varepsilon(i\xi)=1+
\frac{\varepsilon_0-1}{1+\frac{\xi^2}{\omega_e^2}}.
\label{70}
\eeq
\ni
Here $\omega_e\sim 2\times 10^{16}\,$Hz gives the main
electronic absorption in the ultraviolet,
$\varepsilon_0$ is the static dielectric constant. At small
$\xi\ll\omega_e$ one has 
$\varepsilon(i\xi)\approx\varepsilon_0$.
In fact, frequencies giving large contributions to
the Lifshitz formulas (\ref{1}) or (\ref{37}) in the
micrometer separation range are much smaller than
$\omega_e$. Because of this, $\varepsilon(i\xi)$
can be approximately replaced by $\varepsilon_0$.
Below we use in all computations 
$\varepsilon_0\approx 7$ which corresponds to the sheet of mica.

In reality, the zeroth term of Lifshitz formula for
dielectrics is also discontinuous as in the case of
metals. To illustrate this statement we put $l=0$
in (\ref{3}) and obtain the following
values of the reflection coefficients defined in
Eq.~(\ref{2})
\beq
r_1^2(0,k_{\perp})=\left(\frac{\varepsilon_0
-1}{\varepsilon_0+1}\right)^2,
\quad
r_2^2(0,k_{\perp})=0.
\label{71}
\eeq
\ni
These values do not depend on $k_{\perp}$.
Therefore they are preserved in the limit 
$k_{\perp}\to 0$.
At the same time, if we put $k_{\perp}=0$ from the
very beginning one obtains
\beq
r_1^2(\xi_l,0)=r_2^2(\xi_l,0)=r^{(R)}
\equiv
\left(\frac{\sqrt{\varepsilon_0}
-1}{\sqrt{\varepsilon_0}+1}\right)^2,
\label{72}
\eeq
\ni
which is the case for real photons. These values do not depend on
$\xi_l$ and are preserved in the limit $\xi_l\to 0$.
Eqs.~(\ref{71}), (\ref{72}) together imply that both
reflection coefficients $r_1(\xi,k_{\perp})$ and
$r_2(\xi,k_{\perp})$ are discontinuous as the functions
of two variables at a point (0,0). Remind that in the
case of metals described by Drude model only the
transverse reflection coefficient $r_2$ was discontinuous
(see Sec.~II). Note that in the zeroth term of the Lifshitz
formula for dielectrics both reflection coefficients
(\ref{71}) correspond to nonphysical (virtual) photons
with $r_1>r^{(R)}$ and $r_2<r^{(R)}$. At the same
time, for metals the longitudinal reflection coefficient 
at zero frequency takes the physical value 
$r_1=r^{(R)}=1$ in both plasma and Drude models.
As to the perpendicular reflection coefficient at 
zero frequency in the case of metals, it corresponds to
nonphysical photons and takes the values
$r_2=0<r^{(R)}=1$ in Drude model and
$r_2=g(k_{\perp})<r^{(R)}=1$ in the plasma model
(see Eqs.~(\ref{13}), (\ref{15})).

It is of paramount importance that the discontinuity
of both reflection coefficients causes no physical
problem in the cases of dielectrics. The point is that
the dielectric permittivity 
(\ref{70}) corresponds to a nondissipative medium
(which is not the case for metals described by the Drude model).
For this reason, the scattering problem, which
underlies the Lifshitz theory (see Sec.~II and also
\cite{43}), is well defined at zero frequency through the unitarity
of scattering matrix, 
furnishing the desired value of the
scattering coefficient and thereby the free energy.
The results obtained for dielectrics are physically 
consistent. They are immediately evident from
Eqs.~(\ref{1}), (\ref{37}) without use of any additional
assumptions which are necessary in both cases of
ideal and real metals where the scattering problem at
zero frequency is not well defined.

The results of numerical computations for the dielectric
test bodies made of mica are shown by the solid
curves 2 in Fig.~\ref{forceT} (Eq.~(\ref{1}), two plane
plates) and in Fig.~\ref{energyT} (Eq.~(\ref{37}),
a lens above a plate). The curve 2 is in direct analogy with the curve 1
in the same figure. At all separations
the relative temperature correction for dielectrics is
positive. At $a=0.1\,\mu$m it takes the value
$\delta_T(F_{ss})=1.94\times 10^{-5}$. This is larger
than for ideal metal and for real metal ($Al$) considered 
in framework of the plasma model but smaller than for the same
metal in framework of the Drude model (see Table I).
At $a=1\,\mu$m and $a=10\,\mu$m there are
$\delta_T(F_{ss})=1.99\times 10^{-2}$, respectively,
$\delta_T(F_{ss})=2.50$ for dielectrics. At
$a\geq 1\,\mu$m the temperature correction for dielectrics
is larger than for ideal or real metals at the 
same separation. For the configuration of a dielectric
lens above a disk at separations $a=0.1,\>1$ and
$10\,\mu$m the temperature correction is, respectively,
$2.39\times 10^{-4}$, $6.94\times 10^{-2}$, and 4.25.
The relationship of these values with those
computed for the real and ideal metals (see
Table II) is the same as in the case of two plane plates.

\section{CONCLUSIONS AND DISCUSSION}

As demonstrated above, the computation of the Casimir
force between real metals at nonzero temperature is
a complicated theoretical problem. The first contradictions
between Lifshitz theory \cite{2,15} applied to ideal
metals and calculations based on quantum field theory 
\cite{23,24} were revealed in the sixties. They
were resolved by Schwinger, DeRaad and Milton \cite{17}
by the use of special prescription modifying the
zero-frequency term of Lifshitz formula. After the use
of this prescription the results of \cite{2,15} in
application to ideal metal became agreed with those
of \cite{23,24}.

A new stage in the solution of the problem
has been started only recently. It was motivated by the
increased accuracy of the Casimir force measurements
and possible applications of the Casimir effect as a test
of fundamental physical theories and in nanotechnology.
Different authors \cite{27,28,29,30,31,32} applied
Lifshitz theory to calculate the temperature Casimir
force between real metals and obtained the diversed
results. In \cite{29,30} the Lifshitz formula was applied
to real metals in its original form without 
any modification of the zero-frequency term. The obtained
results, however, turned out to be in contradiction
with experiment (see Sec.~V) and with general theoretical
requirements (negative temperature corrections at small
distances and incorrect asymptotic at high temperatures,
see Secs.~II, IV, V). In \cite{31,32} the zeroth term of
Lifshitz formula for real metals at nonzero temperature
was modified according to the prescription of \cite{17}
formulated for ideal metals. The obtained results were
compared with the Casimir force at zero temperature
which was computed for real metals without use of any
prescription. Such an approach leads to significant temperature
corrections to the Casimir force at small separations
(both in plasma and Drude models)
which are linear in temperature and also to the absence of any
finite conductivity corrections at the 
moderate separations. In \cite{27,28} the computations
of the temperature Casimir force in frames of plasma
model were performed with the coinciding results (no 
linear in temperature corrections were found at small 
distances). These computations did not use any modification
of Lifshitz formula. In \cite{28} the problems arising in 
frames of Drude model were also formulated and the way 
to their solution was directed.

In the present paper we propose the new prescription 
modifying the zeroth term of Lifshitz formula in the
case of real metals described by Drude model (Secs.~II,
III). The necessity of this prescription has in-depth
reasons connected with the failure of scattering 
formalism underlying Lifshitz formula in the case when the
dielectric permittivity describes a medium with dissipation
where the unitarity condition is absent. 
This prescription is the generalization
of the Schwinger, DeRaad and Milton prescription \cite{17}
for the case of real metal. In the case of
plasma model (which describes a nondissipative medium) 
it leads to exactly the same results as an
unmodified Lifshitz formula. Because of this all the
results obtained in \cite{27,28} preserve their validity.

The Lifshitz formula with the modified zero-frequency
term is given by Eq.~(\ref{35}) (configuration of two
plane plates) and by Eq.~(\ref{48}) (configuration of a
lens above a plate). The detailed computations with the
use of these equations were performed in Secs.~IV, V.
It was shown that the obtained temperature corrections
are positive and offer the correct asymptotic behavior
both at small and high temperatures (separations).
The results obtained in frames of Drude model join
smoothly with those obtained in frames of plasma model
when the relaxation frequency goes to zero. The finite
conductivity corrections to the Casimir force were computed
at nonzero temperature in the separation range from
$0.1\,\mu$m till $10\,\mu$m in frames of Drude and plasma
models. Both the temperature and finite conductivity
corrections calculated above possess the reasonable
physical properties avoiding difficulties which arise in
\cite{29,30} and in \cite{31,32}. The perturbative
analytical results at both small and large separations
are in agreement with the numerical computations. The case
of dielectric test bodies (Sec.~VI) where the scattering
problem is consistent at all frequencies illustrates
the essence of difficulties arising in the case of
real metals.

In the nearest future one should expect the experimental
registration of the temperature Casimir force. This will
be the final answer in the discussion on the subject
what is the temperature dependence of the Casimir force.
Meanwhile it is necessary to proceed with the more detailed
elaboration of the microscopic theory of dispersion forces
based on quantum field theory at nonzero temperature in
Matsubara formulation.
There is a real possibility that the above phenomenological
prescription, concerning the zero-frequency contribution 
to Lifshitz formula, will be rigorously derived, at least
as a good approximation,
in terms of the scattering theory in dissipative media. 
The final solution of this problem seems to be of
large importance taking into account the prospective role of dispersion
forces in both fundamental and applied science.

\section*{ACKNOWLEDGMENTS}

The authors are grateful for helpful discussions
to M.Bordag,
I.A. Merkulov, U.~Mohideen, and V.I.~Perel'.
 They are indebted to the Brazilian Center of Physical
Research (Rio de Janeiro) and 
Physics Department of the Federal University
of Para\'{\i}ba (Jo\~{a}o Pessoa)
for kind hospitality. This work was
partially supported by FAPERJ and CNPq.

\section*{APPENDIX I}
\setcounter{equation}{0}
\renewcommand{\theequation}{I.\arabic{equation}}

In this Appendix Eqs.~(\ref{26}), (\ref{27}) are derived starting from
Eq.~(\ref{25}). Using (\ref{20}), Eq.~(\ref{25}) can be rewritten in
the form
\bes
&&
F_{ss}(a)=
-\frac{\hbar c}{32\pi^2 a^4}
\int\limits_{0}^{\infty}y^2\,dy
\left\{\frac{1}{2}
\int\limits_{0}^{y}d{\tilde\xi}
f_{ss}(\tilde\xi,y)\right.
\nn \\
&&\phantom{F_{ss}(a)}
+\frac{\pi}{2t}\left[f_{ss}^{(1)}(y,y)-
\int\limits_{0}^{y}d{\tilde\xi}
\frac{\partial f_{ss}^{(1)}(\tilde\xi,y)}{\partial{\tilde\xi}}\right]
\nn \\
&&\phantom{F_{ss}(a)}
+\frac{\pi}{2t}\left[f_{ss}^{(2)}(y,y)-
\int\limits_{0}^{y}d{\tilde\xi}
\frac{\partial f_{ss}^{(2)}(\tilde\xi,y)}{\partial{\tilde\xi}}\right]
\nn \\
&&\phantom{F_{ss}(a)}
-\frac{1}{2}yf_{ss}(y,y)
\nn \\
&&\phantom{F_{ss}(a)}
+\frac{1}{2}\left[{\tilde\xi} f_{ss}(\tilde\xi,y)
{\left.\right\vert}_0^y-
\int\limits_{0}^{y}d{\tilde\xi}f_{ss}(\tilde\xi,y)\right]
\nn \\
&&\phantom{F_{ss}(a)}
+\frac{\pi}{t}f_{ss}(y,y)\,A\left(\frac{ty}{2\pi}\right)
\nn \\
&&\phantom{F_{ss}(a)}
-\frac{\pi}{t}\left.
\int\limits_{0}^{y}d{\tilde\xi}
\frac{\partial f_{ss}(\tilde\xi,y)}{\partial{\tilde\xi}}
A\left(\frac{t{\tilde\xi}}{2\pi}\right)
\right\}.
\label{I1}
\ees

The functions $f_{ss}^{(1)}$ and $\partial f_{ss}^{(1)}/\partial\tilde\xi$
are continuous at all points for both plasma and Drude models. Due to this
it holds
\beq
\int\limits_{0}^{y}d{\tilde\xi}
\frac{\partial f_{ss}^{(1)}(\tilde\xi,y)}{\partial{\tilde\xi}}=
f_{ss}^{(1)}(y,y)-f_{ss}^{(1)}(0,y).
\label{I2}
\eeq
\ni
Note that the same equality is not valid for $f_{ss}^{(2)}$ because it
is discontinuous at zero frequency. Substituting (\ref{I2}) into (\ref{I1})
and performing the cancellations one obtains
\bes
&&
F_{ss}(a)=
-\frac{\hbar c}{16\pi^2 a^4}
\int\limits_{0}^{\infty}y^2\,dy
\left\{
\frac{\pi}{2t}\left[f_{ss}^{(1)}(0,y)+f_{ss}^{(2)}(y,y)\right]
\right.
\nn \\
&&\phantom{F_{ss}(a)}
-\frac{\pi}{2t}
\int\limits_{0}^{y}d{\tilde\xi}
\frac{\partial f_{ss}^{(2)}(\tilde\xi,y)}{\partial{\tilde\xi}}
\nn \\
&&\phantom{F_{ss}(a)}
+\frac{\pi}{t}f_{ss}(y,y)\,A\left(\frac{ty}{2\pi}\right)
\nn \\
&&\phantom{F_{ss}(a)}
-\frac{\pi}{t}\left.
\int\limits_{0}^{y}d{\tilde\xi}
\frac{\partial f_{ss}(\tilde\xi,y)}{\partial{\tilde\xi}}
A\left(\frac{t{\tilde\xi}}{2\pi}\right)
\right\}.
\label{I3}
\ees
\ni
Now let us consider two last contributions to (\ref{I3}) containing the
function $A(z)$. With account of Eq.~(\ref{19})
$\tau=2\pi T/T_{eff}=2\pi/t$. According to definition of the integer
portion function
\beq
A\left(\frac{ty}{2\pi}\right)=A\left(\frac{y}{\tau}\right)=
k{\ }\mbox{if}{\ }k\tau\leq y<(k+1)\tau
\label{I4}
\eeq
\ni
for $k=0,\,1,\,2,\ldots\>$.
Using (\ref{I4}) the integral from (\ref{I3}) can be represented as
\bes
&&
\int\limits_{0}^{\infty}y^2\,dy
f_{ss}(y,y)\,A\left(\frac{ty}{2\pi}\right)=
\int\limits_{\tau}^{2\tau}y^2\,dy\,f_{ss}(y,y)
\nn \\
&&\phantom{(a)}
+2\int\limits_{2\tau}^{3\tau}y^2\,dy\,f_{ss}(y,y)+\cdots +
l\int\limits_{l\tau}^{(l+1)\tau}y^2\,dy\,f_{ss}(y,y)+\cdots
\nn \\
&&\phantom{(a)}
=\int\limits_{\tau}^{\infty}y^2\,dy\,f_{ss}(y,y)+
\int\limits_{2\tau}^{\infty}y^2\,dy\,f_{ss}(y,y)+\cdots+
\int\limits_{l\tau}^{\infty}y^2\,dy\,f_{ss}(y,y)+\cdots
\nn \\
&&\phantom{(a)}
=\sum\limits_{l=1}^{\infty}
\int\limits_{l\tau}^{\infty}y^2\,dy\,f_{ss}(y,y).
\label{I5}
\ees
\ni
The second integral from (\ref{I3})
containing the integer portion function is a bit more
complicated. Using the definition of this function, it can be represented
in the following form
\bes
&&
\int\limits_{0}^{\infty}y^2\,dy
\int\limits_{0}^{y}d{\tilde\xi}
\frac{\partial f_{ss}(\tilde\xi,y)}{\partial{\tilde\xi}}
A\left(\frac{t{\tilde\xi}}{2\pi}\right)=
\int\limits_{\tau}^{2\tau}y^2\,dy
\int\limits_{\tau}^{y}d{\tilde\xi}
\frac{\partial f_{ss}(\tilde\xi,y)}{\partial{\tilde\xi}}
\nn \\
&&\phantom{(a)}
+\int\limits_{2\tau}^{3\tau}y^2\,dy\left[
\int\limits_{\tau}^{2\tau}d{\tilde\xi}
\frac{\partial f_{ss}(\tilde\xi,y)}{\partial{\tilde\xi}}+
2\int\limits_{2\tau}^{y}d{\tilde\xi}
\frac{\partial f_{ss}(\tilde\xi,y)}{\partial{\tilde\xi}}\right]+\cdots
\nn \\
&&\phantom{(a)}
+\int\limits_{l\tau}^{(l+1)\tau}y^2\,dy\left[
\int\limits_{\tau}^{2\tau}d{\tilde\xi}
\frac{\partial f_{ss}(\tilde\xi,y)}{\partial{\tilde\xi}}+
2\int\limits_{2\tau}^{3\tau}d{\tilde\xi}
\frac{\partial f_{ss}(\tilde\xi,y)}{\partial{\tilde\xi}}+\cdots
\right.
\nn \\
&&\phantom{(a)}
\left.
+l\int\limits_{l\tau}^{y}d{\tilde\xi}
\frac{\partial f_{ss}(\tilde\xi,y)}{\partial{\tilde\xi}}\right]+\cdots\, .
\label{I6}
\ees

Now we calculate all integrals with respect to $\tilde\xi$ according to
\beq
\int\limits_{a}^{b}d{\tilde\xi}
\frac{\partial f_{ss}(\tilde\xi,y)}{\partial{\tilde\xi}}=
f_{ss}(b,y)-f_{ss}(a,y).
\label{I7}
\eeq
\ni
For all $a\neq 0$, as it is in (\ref{I6}), Eq.~(\ref{I7}) is valid for both
polarizations (i.e. for both $f_{ss}^{(1)}$ and $f_{ss}^{(2)}$) because
the quantities under consideration are continuous. The result is
\bes
&&
\int\limits_{0}^{\infty}y^2\,dy
\int\limits_{0}^{y}d{\tilde\xi}
\frac{\partial f_{ss}(\tilde\xi,y)}{\partial{\tilde\xi}}
A\left(\frac{t{\tilde\xi}}{2\pi}\right)=
\int\limits_{\tau}^{2\tau}y^2\,dy
\left[f_{ss}(y,y)-f_{ss}(\tau,y)\right]
\nn \\
&&\phantom{(a)}
+\int\limits_{2\tau}^{3\tau}y^2\,dy
\left[2f_{ss}(y,y)-f_{ss}(\tau,y)-f_{ss}(2\tau,y)\right]+\cdots
\label{I8} \\
&&\phantom{(a)}
+\int\limits_{l\tau}^{(l+1)\tau}y^2\,dy
\left[lf_{ss}(y,y)-f_{ss}(\tau,y)-f_{ss}(2\tau,y)
-\cdots - f_{ss}(l\tau,y)\right]+\cdots\, .
\nn
\ees
\ni
Combining the terms with identical arguments and using (\ref{I5}), we
obtain
\bes
&&
\int\limits_{0}^{\infty}y^2\,dy
\int\limits_{0}^{y}d{\tilde\xi}
\frac{\partial f_{ss}(\tilde\xi,y)}{\partial{\tilde\xi}}
A\left(\frac{t{\tilde\xi}}{2\pi}\right)=
-\sum\limits_{l=1}^{\infty}
\int\limits_{l\tau}^{\infty}y^2\,dy\,f_{ss}(l\tau,y)
\nn \\
&&\phantom{aaaaaaaaaaaaaaaa}
+\sum\limits_{l=1}^{\infty}
\int\limits_{l\tau}^{\infty}y^2\,dy\,f_{ss}(y,y).
\label{I9}
\ees

Now we substitute (\ref{I5}) and (\ref{I9}) into (\ref{I3}) with the result
\bes
&&
F_{ss}(a)=
-\frac{\hbar c}{32\pi a^4 t}
\int\limits_{0}^{\infty}y^2\,dy
\left[\vphantom{\int\limits_{0}^{y}}
f_{ss}^{(1)}(0,y)+f_{ss}^{(2)}(y,y)\right.
\nn \\
&&\phantom{aaa}
-\left.
\int\limits_{0}^{y}d{\tilde\xi}
\frac{\partial f_{ss}^{(2)}(\tilde\xi,y)}{\partial{\tilde\xi}}\right]
-\frac{\hbar c}{16\pi a^4 t}
\sum\limits_{l=1}^{\infty}
\int\limits_{l\tau}^{\infty}y^2\,dy\,f_{ss}(l\tau,y).
\label{I10}
\ees
\ni
With account of $t\equiv T_{eff}/T=\hbar c/(2ak_BT)$ and
$l\tau\equiv{\tilde\xi}_l$ Eq.~(\ref{I10}) coincides with Eqs.~(\ref{26}),
(\ref{27}).

\section*{APPENDIX II}
\setcounter{equation}{0}
\renewcommand{\theequation}{II.\arabic{equation}}

In this Appendix Eqs.~(\ref{57}), (\ref{67}) are
obtained and the integral quantities $I_n(\tilde\gamma)$
($i=1,\,2$) are computed. The expression under the second
integral in the right-hand side of Eq.~(\ref{56}) is
approximately equal to
\beq
\frac{y^2}{r_2^{-2}e^y-1}\approx
\frac{y^2}{e^y-1}-
2\frac{\delta_0}{a}y^2\sqrt{y(y+\tilde\gamma)}
\frac{e^y}{(e^y-1)^2},
\label{II1}
\eeq
\ni
where the terms up to the first order in small parameter
$\delta_0/a$ are preserved. Notice that the first 
contribution here is the same as under the first integral
in (\ref{56}). They together produce the high-temperature
asymptotic for the case of ideal metal. The second
contribution in (\ref{II1}) takes the effects of finite
conductivity into account.
This second contribution can be identically represented as
\beq
-2\frac{\delta_0}{a}y^2\sqrt{y(y+\tilde\gamma)}
\frac{e^y}{(e^y-1)^2}=
-2\frac{\delta_0}{a}\left[\frac{y^3e^y}{(e^y-1)^2}+
\tilde\gamma\frac{y^2\sqrt{y}}{\sqrt{y+\tilde\gamma}
+\sqrt{y}}\,\frac{e^y}{(e^y-1)^2}\right].
\label{II2}
\eeq

Substituting (\ref{II1}), (\ref{II2}) with account of
$2\delta_0/a=4/{\tilde\omega}_p$ and 
$\tilde\gamma/{\tilde\omega}_p\equiv\gamma/\omega_p$
into (\ref{56}) one obtains
\beq
F_{ss}(a)=-\frac{k_BT}{16\pi a^3}
\left[4\zeta(3)-12\zeta(3)\frac{\delta_0}{a}-
4\frac{\gamma}{\omega_p}I_1(\tilde\gamma)\right],
\label{II3}
\eeq
\ni
where $I_1(\tilde\gamma)$ was defined in (\ref{58}).
This equation coincides with Eq.~(\ref{57}) because
of (\ref{59}).

\begin{figure}[h]
\epsfxsize=15cm\centerline{\epsffile{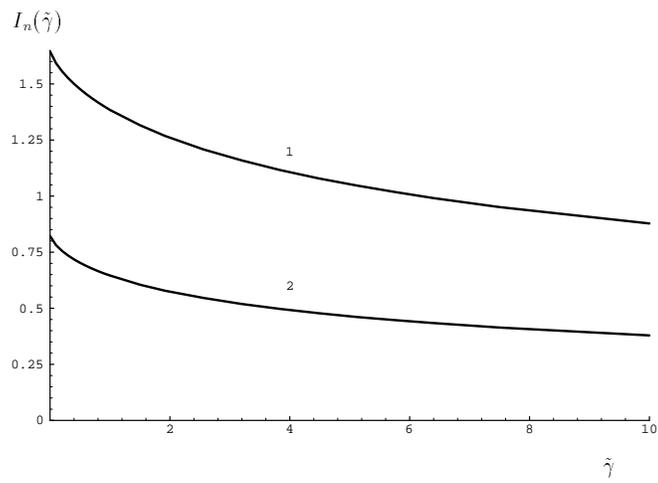}}
\vspace*{-4cm}
\caption{\label{dr-ht}
Dependence of the coefficient integrals in Eqs.~(57)
(curve 1) and (67) (curve 2) on the dimensionless
relaxation frequency.
}
\end{figure}

In Fig.~\ref{dr-ht} the dependence of $I_1$ on 
$\tilde\gamma$ is plotted (curve 1). The values of $I_1$
decrease from 1.3844 at $\tilde\gamma=1$ until 0.8782
at $\tilde\gamma=10$. With account of definition
of Sec.~II $\tilde\gamma=2a\gamma/c$ the dependence
of $I_1$ on $\tilde\gamma$ may be recalculated into the
dependence of $I_1$ on $a$ at fixed $\gamma$. To take an
example, with the above value of 
$\gamma=9.6\times 10^{13}\,$rad/s $\tilde\gamma$ changes
in the interval $1.92\leq\tilde\gamma\leq 6.4$ when
the separation distance belongs to the interval
$3\,\mu\mbox{m}\leq a\leq 10\,\mu$m.

We now turn to the derivation of Eq.~(\ref{67}).
The expression under the second
integral in the right-hand side of Eq.~(\ref{66}) 
calculated up to the first order in $\delta_0/a$ is
\beq
y\,\ln\left[1-r_2^{2}(y,y)e^{-y}\right]\approx
y\,\ln(1-e^{-y})+
2\frac{\delta_0}{a}\sqrt{y(y+\tilde\gamma)}\,
\frac{y}{e^y-1}.
\label{II4}
\eeq
\ni
Once more the first 
contribution here is the same as under the first integral
in (\ref{66}). Together they produce the high-temperature
Casimir force between a lens and a plate
made of ideal metal. The second
contribution in (\ref{II4}) is responsible for the finite
conductivity correction.
This second contribution can be rewritten in the form
\beq
2\frac{\delta_0}{a}\sqrt{y(y+\tilde\gamma)}
\frac{y}{e^y-1}=
2\frac{\delta_0}{a}\left(\frac{y^2}{e^y-1}+
\tilde\gamma\frac{y\sqrt{y}}{\sqrt{y+\tilde\gamma}
+\sqrt{y}}\,\frac{1}{e^y-1}\right).
\label{II5}
\eeq

Substituting (\ref{II4}) and (\ref{II5}) 
into Eq.~(\ref{66}) and performing the integration
one obtains
\beq
F_{sl}(a)=\frac{k_BTR}{8a^2}
\left[-2\zeta(3)+4\zeta(3)\frac{\delta_0}{a}+
4\frac{\gamma}{\omega_p}I_2(\tilde\gamma)\right],
\label{II6}
\eeq
\ni
where $I_2(\tilde\gamma)$ was defined in (\ref{68}).
Eq.~(\ref{II6}) coincides with Eq.~(\ref{67}) if to
take account of (\ref{69}).

In Fig.~\ref{dr-ht} the dependence of $I_2$ on 
$\tilde\gamma$ is shown by the curve 2. 
The values of $I_2$
decrease from 0.6455 at $\tilde\gamma=1$ until 0.3787
at $\tilde\gamma=10$.

\newpage

\begin{table}[h]
\caption{The relative temperature correction 
to the Casimir force between two plates in
dependence of separation for different models of
a metal conductivity}
\begin{tabular}{ccccccc}
{\ }&\multicolumn{2}{c}{Present paper}& Ideal metal
& Approach of \cite{30} 
&\multicolumn{2}{c}{Approach of \cite{31,32}}\\
$a\,(\mu\mbox{m})$ &
$\delta_T(F_{ss}^D)$ & $\delta_T(F_{ss}^p)$ &
$\delta_T(F_{ss}^0)$ & $\delta_T(F_{ss}^D)$ &
$\delta_T(F_{ss}^D)$ & $\delta_T(F_{ss}^p)$ \\ 
0.1 & $5.16\times 10^{-3}$& $6.60\times 10^{-6}$&
$1.57\times 10^{-7}$&$-7.72\times 10^{-3}$&
$2.18\times 10^{-2}$&$1.62\times 10^{-2}$\\
0.3 & $9.24\times 10^{-3}$&$5.48\times 10^{-5}$&
$1.27\times 10^{-5}$&$-3.49\times 10^{-2}$&
$2.54\times 10^{-2}$&$1.53\times 10^{-2}$\\
0.5 & $1.08\times 10^{-2}$&$2.11\times 10^{-5}$&
$9.82\times 10^{-5}$&$-6.43\times 10^{-2}$&
$2.70\times 10^{-2}$&$1.52\times 10^{-2}$\\
0.7 & $1.18\times 10^{-2}$&6$.05\times 10^{-4}$&
$3.77\times 10^{-4}$&$-9.41\times 10^{-2}$&
$2.83\times 10^{-2}$&$1.54\times 10^{-2}$\\
1 & $1.37\times 10^{-2}$&$2.06\times 10^{-3}$&
$1.57\times 10^{-3}$& --0.138&
$3.06\times 10^{-2}$&$1.68\times 10^{-2}$\\
3 & 0.136 & 0.123 & 0.117 & --0.324 & 0.156 & 0.137 \\
5 & 0.580 & 0.563 & 0. 553 & --0.185 & 0.602 & 0.577 \\
7 & 1.17 & 1.15 & 1.14 & $9.75\times 10^{-2}$&
1.19 & 1.16 \\
10 & 2.08 & 2.06 & 2.05 & 0.556 & 2.11 & 2.07
\end{tabular}
\end{table}
\begin{table}[h]
\caption{The relative temperature correction 
to the Casimir force between a lens and a plate in
dependence of separation for different models of
a metal conductivity}
\begin{tabular}{ccccccc}
{\ }&\multicolumn{2}{c}{Present paper}& Ideal metal
& Approach of \cite{30} 
&\multicolumn{2}{c}{Approach of \cite{31,32}}\\
$a\,(\mu\mbox{m})$ &
$\delta_T(F_{sl}^D)$ & $\delta_T(F_{sl}^p)$ &
$\delta_T(F_{sl}^0)$ & $\delta_T(F_{sl}^D)$ &
$\delta_T(F_{sl}^D)$ & $\delta_T(F_{sl}^p)$ \\ 
0.1 & $6.87\times 10^{-3}$& $6.68\times 10^{-5}$&
$3.09\times 10^{-5}$&$-1.51\times 10^{-2}$&
$2.34\times 10^{-2}$&$1.59\times 10^{-2}$\\
0.3 & $1.13\times 10^{-2}$&$1.08\times 10^{-3}$&
$8.09\times 10^{-4}$&$-5.76\times 10^{-2}$&
$2.76\times 10^{-2}$&$1.62\times 10^{-2}$\\
0.5 & $1.56\times 10^{-2}$&$4.33\times 10^{-3}$&
$3.63\times 10^{-3}$&$-9.96\times 10^{-2}$&
$3.22\times 10^{-2}$&$1.92\times 10^{-2}$\\
0.7 & $2.27\times 10^{-2}$&$1.09\times 10^{-2}$&
$9.63\times 10^{-3}$&--0.139&
$3.97\times 10^{-2}$&$2.57\times 10^{-2}$\\
1 & $4.13\times 10^{-2}$&$2.91\times 10^{-2}$&
$2.67\times 10^{-2}$& --0.189&
$5.89\times 10^{-2}$&$4.38\times 10^{-2}$\\
3 & 0.498 & 0.481 & 0.470 & --0.192 & 0.519 & 0.496 \\
5 & 1.33 & 1.31 & 1.30 & 0.183 & 1.36 & 1.32 \\
7 & 2.24 & 2.22 & 2.20 & 0.636& 2.27 & 2.23 \\
10 & 3.62 & 3.58 & 3.57 & 1.33 & 3.65 & 3.60
\end{tabular}
\end{table}

\newpage

\newpage
\newpage
\newpage
\newpage
\newpage
\newpage
\newpage

\end{document}